\documentclass[pdflatex,sn-mathphys-num]{sn-jnl}%
\usepackage{graphicx}%
\usepackage{multirow}%
\usepackage{amsmath,amssymb,amsfonts}%
\usepackage{amsthm}%
\usepackage{mathrsfs}%
\usepackage[title]{appendix}%
\usepackage{xcolor}%
\usepackage{textcomp}%
\usepackage{manyfoot}%
\usepackage{booktabs}%
\usepackage{algorithm}%
\usepackage{algorithmicx}%
\usepackage{algpseudocode}%
\usepackage{listings}%
\usepackage{verbatim}%

\theoremstyle{thmstyleone}%

\theoremstyle{thmstyletwo}%

\theoremstyle{thmstylethree}%

\begin{document}

\title[Fragmentation of $^{20}$Ne]{Modeling fragmentation of $^{20}$Ne nuclei at the LHC}

\author[1,2]{\fnm{Aleksandr} \sur{Svetlichnyi}}\email{aleksandr.svetlichnyy@phystech.edu}

\author[1,2]{\fnm{Savva} \sur{Savenkov}}\email{savenkov.sd@phystech.edu}

\author*[1,2]{\fnm{Igor} \sur{Pshenichnov}}\email{pshenich@inr.ru}

\affil[1]{\orgname{Institute for Nuclear Research of the Russian Academy of Sciences}, \orgaddress{\street{Prospekt 60-letiya Oktyabrya 7a}, \city{Moscow}, \postcode{117312}, \country{Russia}}}

\affil[2]{\orgname{Moscow Institute of Physics and Technology}, \orgaddress{\street{Institutskiy per. 9, Dolgoprudny}, \city{Moscow Region}, \postcode{141701}, \country{Russia}}}

\abstract{Fragmentation of relativistic neon nuclei was studied with the Abrasion–Ablation Monte Carlo for Colliders model (AAMCC-MST), which incorporates pre-equilibrium clustering of spectator matter via a Minimum Spanning Tree (MST) algorithm, subsequent decays of the excited clusters, and Coulomb repulsion between final fragments. After validating the model against data on $^{20}$Ne and $^{22}$Ne fragmentation on various targets, the production of spectator nuclear fragments, neutrons, and protons in $^{20}$Ne--$^{20}$Ne collisions at the LHC was modeled at $\sqrt{s_{\rm NN}}=5.36$~TeV. The production cross sections, transverse and longitudinal momentum distributions for $^2$H, $^4$He, $^6$Li, $^{10}$B, $^{12}$C, $^{14}$N, and $^{16}$O fragments were calculated as these fragments may contaminate $^{20}$Ne beams circulating in the LHC. A simple parametrization describing relative width of the longitudinal momentum distributions calculated for these fragment species was constructed. To study the sensitivity of results to $\alpha$-clustering and nuclear density distributions in neon nuclei, two density profiles for $^{20}$Ne and $^{22}$Ne were used: a deformed Woods–Saxon, and a five–$\alpha$ bi-pyramidal configuration. In addition, a spherical profile was considered for $^{20}$Ne. Comparing results obtained with these profiles and with or without MST clustering shows that the calculated yields of spectator nucleons and fragments, their momentum distributions are sensitive to the details of nuclear structure of $^{20,22}$Ne and fragmentation modeling. The results provide the input for simulations of the transport of secondary forward nuclei in the LHC to estimate their interaction rates with circulating $^{20}$Ne  beam nuclei, and also for  calculations of responses of neutron and proton Zero Degree Calorimeters to $^{20}$Ne--$^{20}$Ne events.}


\maketitle

\section{Introduction}\label{intro}

Collisions of various nuclear systems, namely, pp, $^{129}$Xe--$^{129}$Xe, p--$^{208}$Pb, and $^{208}$Pb--$^{208}$Pb in the multi-TeV center-of-mass energy range were studied during Runs 1 and 2 (2009–2018) at the Large Hadron Collider (LHC) in A Large Ion Collider Experiment (ALICE)~\cite{Acharya2024}, as well as in other LHC experiments. In Run 3 (2022–2026), studies of pp, p--$^{208}$Pb, and $^{208}$Pb--$^{208}$Pb collisions have been extended to higher luminosities using the upgraded ALICE detectors~\cite{AcharyaUpgrade2024}. Furthermore, for the first time at the LHC, light nuclei were accelerated~\cite{AlemanyFernandez2025,Bruce2026} and data for p--$^{16}$O, $^{16}$O--$^{16}$O, and $^{20}$Ne--$^{20}$Ne collisions were collected in July 2025.

The interest in studying collisions of light nuclei at the LHC is driven by several theoretical predictions of novel effects. Among them are the possible emergence of collective flow in small collision systems~\cite{Giacalone2025}, the influence of $\alpha$-clustering in $^{16}$O and $^{20}$Ne on the initial eccentricity and the resulting anisotropic flow in such collisions~\cite{Lim2019}, and the impact of nucleon–nucleon correlations in $^{16}$O on elliptic flow and on fluctuations of the mean transverse momentum of the produced particles~\cite{Zhang2025}. Recent data on collisions of light systems have already provided a wide range of observables that can be confronted with models of strong interactions, quantum chromodynamics, and the quark–gluon plasma~\cite{Modak2026,ATLAS2025,Hayrapetyan2026}.

Relativistic $^{20}$Ne--$^{20}$Ne collisions are of particular interest~\cite{Liu2025}, as they provide an opportunity to search for signatures of the bowling-pin deformation of $^{20}$Ne, which is associated with its bi-pyramidal $\alpha$-clustered configuration ($^4$He+$^{16}$O)~\cite{Bijker2021,Yamaguchi2023}. It was proposed~\cite{LiPei2026} to distinguish  alternative $5\alpha$ and $\alpha$+$^{16}$O cluster configurations in $^{20}$Ne by measuring flow correlation observables in ultracentral collisions of neon nuclei. A comparative study of $^{20}$Ne--$^{20}$Ne and $^{16}$O--$^{16}$O collisions, with particular emphasis on the elliptic flow $v_2\{2\}$ ratio in the 1\% most central events, is especially informative, as it can reveal the characteristic bowling-pin shape of $^{20}$Ne~\cite{Giacalone2025}. In this approach, the relevant experimental systematic uncertainties are substantially reduced by examining the flow ratio of the two collision systems within the same experimental setup.

It is worth noting that, in many studies, simulations of $^{16}$O--$^{16}$O and $^{20}$Ne--$^{20}$Ne collisions primarily concentrate on particle production within the hot and dense overlap region of the colliding nuclei. Considerably less emphasis has been placed on the nuclear matter located outside this overlap zone. This spectator matter remains comparatively cold and consists of free spectator nucleons and nuclear fragments. For example, in Ref.~\cite{Nijs2025}, $^{16}$O--$^{16}$O and $^{20}$Ne--$^{20}$Ne collisions at the LHC were simulated using the Trajectum framework in combination with the GEMINI evaporation model to determine the cross sections for the formation of different secondary spectator nuclei.

At the LHC, spectator nucleons are emitted at very small angles relative to the beam direction and can therefore be detected only with forward hadronic Zero Degree Calorimeters (ZDC) installed sufficiently far from the interaction point (IP), in particular, in ALICE~\cite{Zugravel2025} and ATLAS~\cite{ATLASZDC2025} experiments. A dedicated ATLAS setup has recently been employed~\cite{Sobien2026} to detect forward nuclear fragments produced in p--$^{16}$O, $^{16}$O--$^{16}$O, and $^{20}$Ne--$^{20}$Ne collisions.

The yields of stable spectator nuclei with $2Z = A$, namely $^2$H, $^4$He, $^6$Li, $^{10}$B, $^{12}$C, and $^{14}$N, produced in $^{16}$O--$^{16}$O collisions at the LHC are of special interest because their charge-to-mass ratios are close to that of the $^{16}$O beam nuclei. The calculations presented in Ref.~\cite{Svetlichnyi2023} show that the probability of producing at least one of these nuclei in a $^{16}$O--$^{16}$O collision is nearly 100\%, and that the corresponding inclusive cross section for production of these nuclei is approximately equal to the total hadronic cross section of 1.34~b. These secondary $2Z=A$ nuclei propagate through the LHC magnetic field alongside the $^{16}$O beam and may interact with $^{16}$O nuclei at the interaction points of the LHC experiments, thereby introducing a contamination in the $^{16}$O--$^{16}$O collision samples. As noted in Refs.~\cite{Svetlichnyi2022,Svetlichnyi2023}, making reliable predictions for the yields of $2Z=A$ spectator fragments in $^{16}$O–$^{16}$O collisions is challenging because these yields are sensitive to the structure of the colliding $^{16}$O nuclei, particularly on whether $\alpha$-clustering is present in $^{16}$O.

In $^{20}$Ne--$^{20}$Ne collisions at the LHC, one anticipates the production of a wider range of $2Z = A$ spectator nuclei. Besides the stable $2Z = A$ spectator nuclei already mentioned for $^{16}$O--$^{16}$O collisions, this set additionally comprises stable $^{16}$O and long-lived $^{18}$F. Several studies ~\cite{Bijker2021,Yamaguchi2023,LiPei2026} have shown that clustered configurations make substantial contributions to the ground state of $^{20}$Ne. Consequently, a precise description of $^{20}$Ne--$^{20}$Ne collisions at the LHC requires a robust model that explicitly accounts for the cluster structure of $^{20}$Ne when calculating the yields of spectator nucleons and fragments, as well as their transverse and longitudinal momenta. Before being applied in this context, such a model should first be tested against existing data on $^{20}$Ne fragmentation at lower beam energies~\cite{Zhang2012,Zeitlin2011}. Furthermore, fragmentation data for $^{22}$Ne projectiles, containing two more neutrons than $^{20}$Ne, in nuclear emulsion at higher energies~\cite{Abd-Allah2001,Neagu2013} should also be used to test the validity of the model.

In the present work, we begin by validating the AAMCC-MST model using existing data on the fragmentation of neon nuclei~\cite{Abd-Allah2001,Neagu2013,Zhang2012,Zeitlin2011}. We then calculate a comprehensive set of observables describing the production of spectator neutrons, protons, and fragments in $^{20}$Ne--$^{20}$Ne collisions at $\sqrt{s_{\rm NN}} = 5.36$~TeV. Our analysis concentrates on the production of $2Z=A$ fragments capable of propagating along the $^{20}$Ne beams in the LHC, as well as on the production of spectator neutrons and protons which can be detected by ZDC and other forward detectors installed at the LHC. We also assess how the deformation of the $^{20}$Ne nucleus and the $\alpha$-cluster structure of its ground state influence the results. The  sensitivity of results to the inclusion of pre-equilibrium clusterization of spectator matter is also evaluated. 

\section{Modeling fragmentation of \texorpdfstring{$^{20}$Ne}{20Ne} and \texorpdfstring{$^{22}$Ne}{22Ne} with AAMCC-MST}\label{model}

The main concept of the AAMCC-MST model, its implementation and modeling results for collisions of relativistic nuclei have been described in detail in our papers~\cite{Svetlichnyi2020,Nepeivoda2022}. In Refs.~\cite{Svetlichnyi2022,Svetlichnyi2023} we focused on modeling fragmentation of relativistic oxygen nuclei in fixed target and collider experiments at the LHC. Thus, mostly the main points of modeling are recapped here, with emphasis on the simulation details that are specific to fragmentation of $^{20}$Ne and $^{22}$Ne nuclei. Each nucleus-nucleus collision event is represented by several stages, which are modeled as outlined below.

\subsection{Initial stage of a collision event}\label{initial}

In the AAMCC-MST model, the first initial stage of each nucleus-nucleus collision event is simulated by means of the Glauber Monte Carlo (GlauberMC) model~\cite{Loizides2018}. The positions of nucleons in initial nuclei are sampled according to the nuclear density distributions in $^{20}$Ne and $^{22}$Ne in their ground states. These distributions were parameterized by a spherical Woods-Saxon distribution (hereafter denoted as WS-spherical), a deformed Woods-Saxon distribution (WS-deformed), or explicitly assuming $\alpha$-clustered configurations in these nuclei represented by bi-pyramids (Bi-pyramid)~\cite{Bijker2021,LiPei2026}. The parameters of these distributions are listed in Table~\ref{tab:neon_all_params}. 

The WS-deformed nuclear density distributions of $^{20}$Ne and $^{22}$Ne as functions of the coordinates $x$, $y$, and $z$ were defined as
\begin{equation}
\rho(x,y,z)=\frac{\rho_0}{1+\exp\left\lfloor (r-R(1 + \beta_2 Y_{20} + \beta_4 Y_{40}))/a\right\rfloor}\ ,  
\label{2pF_1}
\end{equation}
where $R$ denotes the half-density nuclear radius, $a$ is the diffuseness parameter, and $r^2 = x^2 + y^2 + z^2$. The quadrupole ($\beta_2$) and hexadecapole ($\beta_4$) even-multipole deformation parameters describe the deviation of the neon nuclei from the purely spherical shape that follows from the distribution given by Eq.~(\ref{2pF_1}) with $\beta_2=\beta_4=0$ corresponding to a simple shell model of $^{20}$Ne. In Eq.~(\ref{2pF_1}), the deformation is expressed via the spherical harmonics $Y_{20}$ and $Y_{40}$:
\begin{equation}
Y_{20} (x,y,z) = \sqrt{\frac{5}{\pi}}\frac{2z^2 - x^2 - y^2}{4r^2}, \  \   \   Y_{40} (x,y,z) = \frac{3}{\sqrt{\pi}}\frac{35z^4 - 30z^2r^2 + 3r^4}{16r^4} \ .
\label{2pF_2}
\end{equation}

In this work, the parameters $R$ and $a$ of the radial nuclear density distribution in Eq.~(\ref{2pF_1}) for $^{20}$Ne and $^{22}$Ne are taken from Refs.~\cite{DeVries1987} and~\cite{Li2023}, respectively. More specifically, these parameter sets correspond to nuclear charge density distributions determined from electron-scattering data. For the $2Z=A$ isospin-symmetric nucleus $^{20}$Ne, one may reasonably assume that $R=2.805$~fm and $a=0.571$~fm~\cite{DeVries1987} apply equally to the neutron density and, consequently, to the overall matter distribution, implying the absence of a noticeable neutron or a proton skin. This expectation is supported by independent measurements of interaction cross sections and matter radii for the $A=20$ isobars~\cite{Chulkov1995}, including $^{20}$Ne, which yield $R=2.816$~fm and $a=0.512$~fm for this nucleus.  As seen from Table~\ref{tab:neon_all_params}, the values of $R$ and $a$ are larger for $^{22}$Ne in comparison to those for $^{20}$Ne because of two additional neutrons in $^{22}$Ne.  
The quadrupole deformation parameter $\beta_{2}$ for $^{20}$Ne is taken from Ref.~\cite{Pritychenko2016}, under the assumption that the hexadecapole deformation is negligible, i.e., $\beta_{4}=0$, while both $\beta_{2}$ and $\beta_{4}$ for $^{22}$Ne are adopted following Ref.~\cite{Moller1995}. To account for arbitrary initial orientations of the colliding nuclei, the orientation of the coordinate system employed in Eqs.~(\ref{2pF_1}) and (\ref{2pF_2}) was randomly sampled in the simulations.

\begin{figure}
    \centering
    \includegraphics[width=0.2\linewidth]{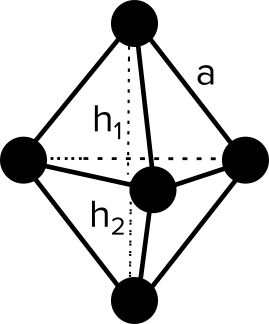}
    \caption{The schematic representation of the bi-pyramid structure of Ne.}
    \label{fig:bi-pyramid}
\end{figure}

In order to model the clustered structure of $^{20}$Ne, a bi-pyramidal configuration~\cite{Bijker2021,LiPei2026} (see Fig.~\ref{fig:bi-pyramid}) was employed. Within this approach, the five $\alpha$ clusters of $^{20}$Ne are taken to form a bi-pyramid. Four of the clusters constitute a $^{16}$O-like core arranged as a regular tetrahedron of height $h_1$ and edge length $\mathsf{a}$, with the clusters placed at the tetrahedron vertices. The separation from the center of the tetrahedral base to the fifth cluster is denoted as $h_2$. In Ref.~\cite{LiPei2026}, to quantify the size and geometry of the clustered configuration, $D_1$ was introduced as the distance from each of the four core clusters to the tetrahedron center, and $D_2$ as the distance from the fifth cluster to the tetrahedron center. The configuration becomes a symmetric bi-pyramid when $D_2/D_1=5/3$.   

Following Ref.~\cite{LiPei2026}, one may choose $D_1=2.0$~fm and $D_2=3.23$~fm to reproduce the root-mean-square (RMS) radius ${\langle r^2\rangle}^{1/2} \approx 3.0$~fm of $^{20,22}$Ne, and also make the bi-pyramid cluster configurations almost symmetric for these nuclei. With the parameters defined in Fig.~\ref{fig:bi-pyramid}, we set the tetrahedral edge length to $\mathsf{a}=3.26$~fm, corresponding to a height $h_1=\sqrt{2/3}\cdot \mathsf{a}=2.66$~fm. The distance between the fifth $\alpha$ cluster and the center of the regular-tetrahedron base was chosen as $h_2 = 2.56$~fm. With these $h_1$ and $h_2$ values, over $10^5$ nucleon configurations with random orientations of deformed $^{20}$Ne were generated to prove that both the RMS radius of $^{20}$Ne and its quadrupole deformation parameter $\beta_2 = 0.721$ were reproduced on average.

The nuclear density distribution of each $\alpha$ cluster was modeled using a harmonic oscillator (HO) parameterization $\rho (r_\alpha) \propto \exp{(-r^2_\alpha/R^2_{\mathrm{c}})}$. The value
$R_{\mathrm{c}} = 1.368$~fm was taken to reproduce the RMS  charge radius of $^4$He,
$\langle r^2_\alpha \rangle ^{1/2} = 1.676$~fm~\cite{DeVries1987} and ensure the  consistency with the selected edge length, $\mathsf{a}\sim 2\times \langle r^2_\alpha \rangle ^{1/2} $, which defines the general scale of the tetrahedron cluster configuration of $^{20}$Ne.

\begin{table}[htb!]
    \centering
    \caption{Parameters of the spherical, deformed Woods-Saxon nuclear density distributions and clustered bi-pyramid density distributions used in the present work for $^{20}$Ne and $^{22}$Ne nuclei.}
    \begin{tabular}{ccccccccc}
    \toprule
      & \multicolumn{2}{c}{WS-spherical} & \multicolumn{4}{c}{WS-deformed} & \multicolumn{2}{c}{Bi-pyramid} \\
      \cmidrule{2-9}
      & $R$, fm & $a$, fm & $R$, fm & $a$, fm & $\beta_2$ & $\beta_4$ & $h_1$, fm & $h_2$, fm \\
      \midrule
      $^{20}$Ne & 2.805\footnotemark[1] & 0.571\footnotemark[1] & 2.805\footnotemark[1] & 0.571\footnotemark[1] & 0.721\footnotemark[2] & 0 & 2.66 & 2.56 \\
      $^{22}$Ne & -- & -- & 2.895\footnotemark[3] & 0.598\footnotemark[3]  & 0.326\footnotemark[4] & 0.225\footnotemark[4] & 2.66 & 2.56 \\
      \botrule
    \end{tabular}
    \label{tab:neon_all_params}
    \footnotetext[1]{Ref.~\cite{DeVries1987}, page 504}
    \footnotetext[2]{Ref.~\cite{Pritychenko2016}, page 54}
     \footnotetext[3]{Ref.~\cite{Li2023}, page 2}
     \footnotetext[4]{Ref.~\cite{Moller1995}, page 222}
\end{table}

In order to reproduce a similar clustered structure of $^{22}$Ne, it was assumed that two extra neutrons with respect to the above-described bi-pyramid clustered configuration in $^{20}$Ne are distributed according to the deformed Woods-Saxon parametrization of $^{22}$Ne. The parameters of the clustered bi-pyramid density distributions used in the present work for $^{20}$Ne and $^{22}$Ne nuclei are summarized in Table~\ref{tab:neon_all_params}. 

The parameters listed in Table~\ref{tab:neon_all_params} are used to sample the positions of nucleons in both colliding nuclei before each collision event generated with the GlauberMC model~\cite{Loizides2018,Loizides2025}. In this model independent binary collisions between nucleons from one nucleus and nucleons from the other are modeled by propagating the nucleons along straight-line trajectories parallel to the beam direction. The probability of a nucleon–nucleon interaction to occur is calculated following Ref.~\cite{Loizides2025} as 
\begin{equation}
    p(b_{\mathrm{NN}}) \propto \frac{\Gamma\left(1/\omega, \frac{\pi b_{\mathrm{NN}}^2}{\sigma_{\mathrm{NN}}^{\mathrm{inel}} \omega} \right)}{\Gamma(1/\omega)}.
    \label{eq:collisionp}
\end{equation}
\noindent Here $\Gamma(z)$ and $\Gamma( \alpha, z)$ are complete and incomplete Euler gamma-functions, respectively, $b_{\mathrm{NN}}$ is the transverse distance between colliding nucleons in the plane orthogonal to the beam axis, $\sigma^{\mathrm{inel}}_{\mathrm{NN}}$ is the total inelastic nucleon–nucleon cross section, and $\omega$ is the parameter of the nucleons overlap profile tuned to reproduce LHC data. Following Ref.~\cite{Loizides2025},  the values  $\sigma^{\mathrm{inel}}_{\mathrm{NN}}=68$~mb and $\omega = 0.4$ were used in modeling $^{20}$Ne--$^{20}$Ne collisions at $\sqrt{s_{\rm NN}} = 5.36$~TeV.

\subsection{Prefragment excitation energy}\label{excitation}

After all nucleon–nucleon interactions in a nucleus–nucleus collision have been simulated with the GlauberMC model, and the involved nucleons from the projectile and target have been labeled as participants, some nucleons from the incoming nuclei may still not experience any collision. These non-interacting nucleons are identified as spectator neutrons and protons, representing the remnants of the original nuclei that remain relatively cold after the collision. The AAMCC-MST model focuses exclusively on the later evolution of this spectator matter, referred to as prefragments~\cite{Nepeivoda2022}. It neither follows the abraded participant nucleons nor the particles produced by nucleon–nucleon interactions in the fireball. In addition, it ignores any interactions between the fireball products and the spectator matter. For nucleus–nucleus collisions at the LHC, this sharp division between participant and spectator regions is well justified, because the spectators continue forward at nearly the speed of light and thus stay close to the fireball only for a very short time interval.

The total number of spectator nucleons in an event determines the mass number of the prefragment, $A_{\mathrm{pf}}$, which represents the remnant of the initial nucleus with its mass number $A_0$. The second ablation stage in the simulation of a nucleus–nucleus collision starts with evaluating the excitation energy of the spectator prefragments originating from each of the two colliding nuclei. In the AAMCC-MST description of collisions of light nuclei~\cite{Svetlichnyi2022,Svetlichnyi2023}, the mean excitation energy per nucleon in a prefragment, $\varepsilon$, is assumed to follow a parabolic dependence. This specific mean excitation energy is calculated as $\varepsilon = \varepsilon_0 \sqrt{1 - \alpha}$, where $\alpha = A_{\mathrm{pf}}/A_0$ is the relative mass of the prefragment, and $\varepsilon_0$ is a model parameter. This phenomenological form was introduced by the ALADIN Collaboration~\cite{Botvina1995}, and within the AAMCC-MST model it was used as part of a hybrid approach to calculate $\varepsilon$ in modeling fragmentation of medium-weight and heavy nuclei in events with a large number of abraded nucleons; see Ref.~\cite{Nepeivoda2022} for further details. 

In AAMCC-MST modeling of 3.7$A$ $^{16}$O fragmentation in nuclear emulsion~\cite{Svetlichnyi2022}, the parabolic dependence $\varepsilon = \varepsilon_0 \sqrt{1 - \alpha}$ was adopted, and the options $\varepsilon_0 = 4$~MeV and $\varepsilon_0 = 8$~MeV were considered. The choice of $\varepsilon_0 = 4$~MeV yielded an adequate agreement with experimental data on oxygen fragmentation~\cite{Svetlichnyi2022}. Hence, in the present study we also use $\varepsilon_0 = 4$~MeV in modeling fragmentation of $^{20,22}$Ne.

\subsection{Prefragment momentum}\label{prefag_mom}

Following a seminal paper on nuclear fragmentation~\cite{Goldhaber1974}, one can estimate the distribution of the total momentum $p^\star$ of a spectator prefragment with the mass number $A_{\mathrm{pf}}$ in the rest frame of the initial nucleus with the mass number $A_0$. It was assumed that such a distribution is described by a Gaussian, $\exp(-{p^\star}^2/2\sigma^2)$, with a characteristic width of the distribution described by a parabolic dependence on $A_{\mathrm{pf}}$:   
\begin{equation}
    \sigma^2 = \sigma_0^2\frac{A_{\mathrm{pf}}(A_0-A_{\mathrm{pf}})}{A_0-1} \ ,
    \label{eq:Goldhaber_sigma}
\end{equation}
with $\sigma_0$ = 90--100~MeV/$c$.  A theoretical estimation on the basis of the Fermi momentum $p_F$ in the initial nucleus,  $\sigma_0^2=\langle {p^\star}^2 \rangle/3 = p^2_{\mathrm{F}}/5$, has been obtained in Ref.~\cite{Goldhaber1974} by taking randomly a nucleon out of the initial nucleus and considering its intranuclear Fermi motion.
With a typical value $p_{\mathrm{F}}$ = 230~MeV/$c$, the value of $\sigma_0 \sim$ 100~MeV/$c$ is confirmed, see also Ref.~\cite{Ma2002} which reports $\sigma_0 \sim$~112--116~MeV/$c$. A larger value $\sigma_0 = 193$~MeV/$c$ has been suggested in Ref.~\cite{Vasyagina2025} following the comparison of calculated with AAMCC-MST and measured transverse momentum distributions of helium fragments produced in fragmentation of gold nuclei in nuclear emulsion. This value of $\sigma_0$ was used in the present work to sample the momentum distribution of prefragments.

\subsection{Pre-equilibrium clustering of spectator matter}\label{mst}

As noted in Ref.~\cite{Nepeivoda2022}, the spectator matter produced after the ablation phase should not, in general, be treated as a single thermodynamic system that has reached equilibrium by the time it decays. In particular, for central or semi-central collisions of nuclei, the spectator region may acquire a narrow, crescent-like geometry with weak internal connectivity due to its low nucleon density. In such events, it is reasonable to expect that spectator matter will fragment into clusters through pre-equilibrium dynamics, without ever attaining full thermodynamic equilibrium across the whole prefragment. The AAMCC-MST model incorporates the pre-equilibrium breakup of spectator prefragments into lighter nuclear clusters by applying the minimum spanning tree (MST) clusterization procedure~\cite{Nepeivoda2022}.  

In this procedure, a minimum spanning tree~\cite{prim1957shortest} is constructed from the 3D coordinates of all prefragment nucleons. They are treated as vertices of a complete, weighted, undirected graph, where each edge weight corresponds to the absolute value of the distance between the two nucleons. The Kruskal algorithm~\cite{cormen2009introduction} is then used to obtain the MST with the smallest possible total edge weight. Next, all "heavy" edges with lengths exceeding the critical nucleon–nucleon separation distance $d$ are removed. The nucleons that remain mutually connected after this pruning step are subsequently identified as excited clusters. It is assumed that the excitation energy of the spectator prefragment is shared between the resulting clusters proportionally to the number of nucleons they contain. In this way, the excitation energy of each cluster on average corresponds to the specific excitation energy of the entire prefragment. 

As noted in Ref.~\cite{Svetlichnyi2022} and the works cited therein, a hot nuclear system that has attained thermal equilibrium is generally already expanded by the time its multifragmentation occurs. Depending on its excitation energy, such a system typically has a density of about 1/4--1/2 of the normal nuclear density. By the same reasoning, one can anticipate that a spectator prefragment undergoes a comparable expansion already during the early, pre-equilibrium stage of its evolution, implying that a more strongly excited system is more dilute when pre-equilibrium clustering occurs. Since Glauber-type models cannot simulate this expansion explicitly, a phenomenological dependence of the parameter $d$ on $\varepsilon$ is implemented in AAMCC-MST. Making $d$ smaller as $\varepsilon$ grows effectively reproduces stronger clusterization in an expanded, hot system. In this work, the parameters defining the $d(\varepsilon)$ dependence were adopted from Ref.~\cite{Vasyagina2025}.

\subsection{Decays of excited spectator fragments}\label{decays_excited}

In AAMCC-MST, statistical models from the Geant4~\cite{allison2016recent} toolkit are generally employed to simulate the decays of excited clusters formed at the pre-equilibrium stage. The selection of the most suitable model depends on the mass and excitation energy of the excited cluster. For $\varepsilon\leq 4$~MeV, the Weisskopf–Ewing model~\cite{Weisskopf1937} is applied to describe sequential nucleon evaporation. At higher excitation energies, the Statistical Multifragmentation Model (SMM)~\cite{bondorf1995statistical} is used instead. A variant of the Fermi Break-Up model~\cite{Fermi1950} is utilized to simulate decays of light excited nuclei up to $^{18}$O for all excitation energies, and this is the primary model employed in the present study to describe the fragmentation of $^{20,22}$Ne. In practice, the evaporation model is activated mostly in very peripheral collisions of $^{20,22}$Ne with target nuclei, in which fewer than 2–4 nucleons are abraded from the initial nuclei. The Coulomb repulsion between the final-state fragments is also taken into account, see Ref.~\cite{Vasyagina2025} for details.  The momenta of the products of decays of excited prefragments and also clusters resulting from MST-clustering procedure are generated following the energy and momentum conservation laws. Finally, at the end of simulation, the momenta of all spectator nucleons and fragments are transformed from the rest frame of the initial $^{20}$Ne nucleus to the laboratory system.  

\section{AAMCC-MST validation with data on fragmentation of \texorpdfstring{$^{20}$Ne}{20Ne} and \texorpdfstring{$^{22}$Ne}{22Ne} }\label{validation}

The key assumptions of the AAMCC-MST model, as well as the particular parameter settings adopted to simulate fragmentation of light nuclei, can be tested by confronting AAMCC-MST results with experimental data on $^{20,22}$Ne fragmentation in collisions with fixed targets~\cite{Abd-Allah2001,Neagu2013,Zhang2012,Zeitlin2011}. While direct measurements of the production of nuclei with $2Z = A$ in neon fragmentation are not currently available, the published data on yields of certain elements with a given charge $Z$ can nevertheless be used to validate the model.

\begin{figure}[tb!]
\centering
\includegraphics[width=0.85\textwidth]{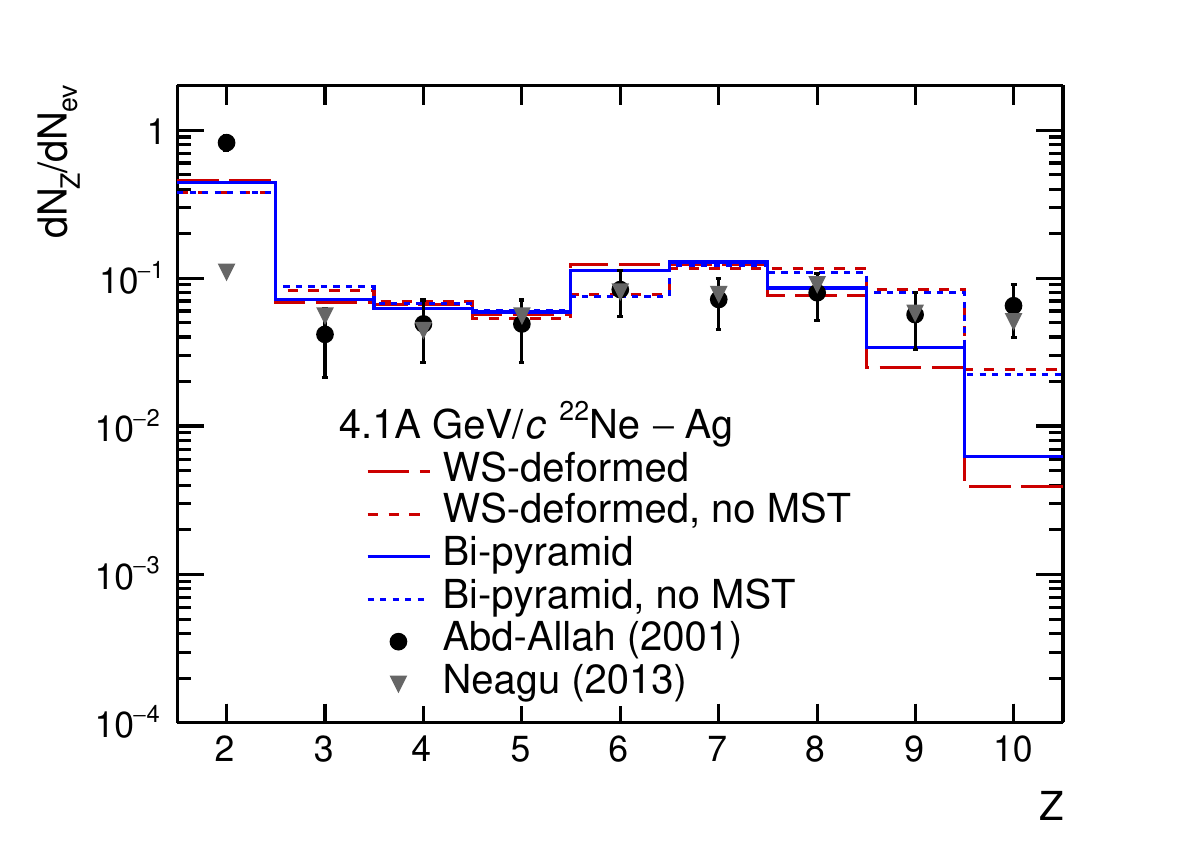}
\caption{Calculated charge distributions of spectator fragments produced in interactions of 4.1$A$~GeV/$c$ $^{22}$Ne projectiles with Ag nuclei. AAMCC-MST results obtained with various calculation options are presented by histograms. Experimental data~\cite{Abd-Allah2001,Neagu2013} on fragmentation of $^{22}$Ne in nuclear emulsion are presented, respectively, by solid circles and triangles.}
\label{fig:charge_0.9}
\end{figure}

Calculated charge distributions, $dN_Z/dN_{\mathrm{ev}}$, for spectator fragments produced in collisions of 4.1$A$~GeV/$c$ $^{22}$Ne projectiles with Ag nuclei are presented in Fig.~\ref{fig:charge_0.9}. Here, $N_Z$ denotes the total number of nuclear fragments of charge $Z$ (i.e., of a given element) generated in the simulations, and $N_{\mathrm{ev}}$ is the total number of simulated events. Silver is taken as a target nucleus representing heavy elements (Ag and Br) in nuclear emulsion. The calculations were carried out using both the WS-deformed and the clustered Bi-pyramid representations of the $^{22}$Ne nuclear density introduced in Sec.~\ref{initial}. For each of these density parameterizations, simulations were performed with and without involving pre-equilibrium MST-clustering. 

Nuclear emulsions were exposed to 4.1$A$~GeV/$c$ $^{22}$Ne beams at the Dubna synchrophasotron~\cite{Abd-Allah2001,Neagu2013}. Events containing projectile fragments were selected, and the ionization density along the fragment tracks was measured to determine the fragment charges $Z$. From these data, the ratios $\sigma(Z)/\sigma_{\mathrm{inel}}$ were obtained, where $\sigma(Z)$ is the cross section for producing a fragment of charge $Z$ and $\sigma_{\mathrm{inel}}$ is the total inelastic reaction cross section. By construction, the simulated quantity $dN_Z/dN_{\mathrm{ev}}$ is equivalent  to the experimental ratio $\sigma(Z)/\sigma_{\mathrm{inel}}$. Results of measurements with the same beam nuclei were reported also in Ref.~\cite{Neagu2013}, but with thorough separation of electromagnetic dissociation (EMD) events in nuclear emulsion from events of fragmentation in hadronic interactions of $^{22}$Ne. The data of Refs.~\cite{Abd-Allah2001,Neagu2013} are presented in Fig.~\ref{fig:charge_0.9}. As can be seen, the outcomes of these two independent experiments are in good mutual agreement excluding the yields of helium fragments, which nevertheless dominate in both measurements.  
 
As shown in Fig.~\ref{fig:charge_0.9}, taking into account measurement uncertainties, our calculations with all four calculation options reproduce well the data~\cite{Abd-Allah2001,Neagu2013} on the production of projectile fragments with charges $Z=3,4,5,6$, and $Z=8,9$, i.e., lithium, beryllium, boron, carbon, oxygen, and fluorine. In contrast, the model overestimates the yield of $Z=7$ nitrogen fragments. It is also evident that the production of $Z=10$ neon fragments is significantly underestimated by our model for all calculation options when compared to the experimental data~\cite{Abd-Allah2001,Neagu2013}. This discrepancy can likely be attributed to the difficulty in separating in the experiments the primary neon beam nuclei from secondary neon nuclei that have lost very few neutrons in peripheral hadronic interactions as well as in EMD events. In other words, a contamination of primary $^{22}$Ne in the $Z=10$ fragment data cannot be ruled out.

As also can be seen in Fig.~\ref{fig:charge_0.9}, the results obtained using the WS-deformed and Bi-pyramid parameterizations of the $^{22}$Ne nuclear density distribution are very similar in general. The calculated fragment yields are more strongly affected by whether MST-clusterization is included in the simulations, particularly for the production of $Z=9,10$ spectator fragments. These relatively heavy fragments are predominantly generated in peripheral collisions and, in calculations without MST-clustering, they are less likely to break up into lighter fragments.

\begin{figure}[tb!]
\centering
\includegraphics[width=0.85\textwidth]{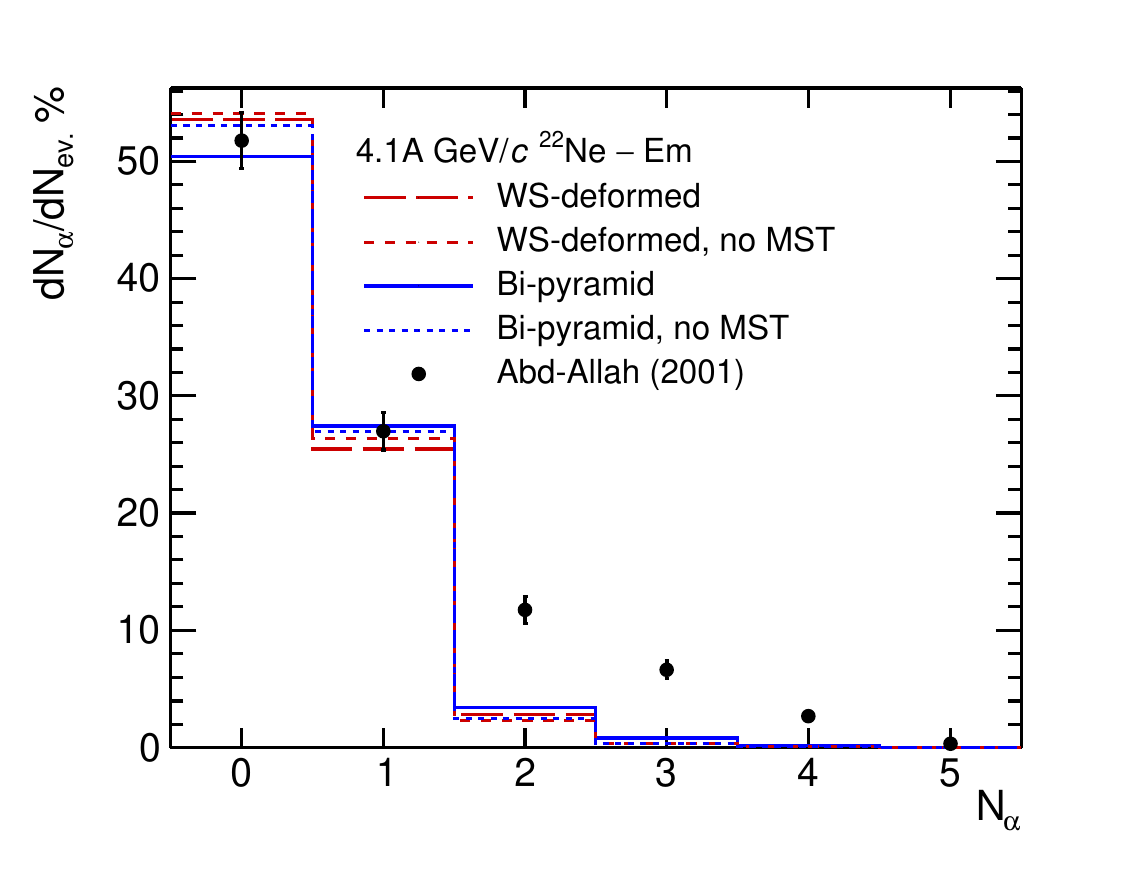}
\caption{Multiplicity distributions of spectator $\alpha$ particles produced in fragmentation of 4.1$A$~GeV/$c$ $^{22}$Ne nuclei in nuclear emulsion. AAMCC-MST results obtained with various calculation options are presented by histograms. Experimental data~\cite{Abd-Allah2001} are presented  by points.}
\label{fig:mult_alphas_comp}
\end{figure}

The calculated yield of helium fragments with $Z=2$ was found to lie between the values reported in Refs.~\cite{Abd-Allah2001} and \cite{Neagu2013}, see Fig.~\ref{fig:charge_0.9}. Since the helium fragments considered as $\alpha$ particles in Refs.~\cite{Abd-Allah2001,Neagu2013} constitute the dominant share of the fragmentation products, it is instructive to compare the $\alpha$-particle multiplicity distributions obtained with the four model options to the experimental results of Ref.~\cite{Abd-Allah2001}, as shown in Fig.~\ref{fig:mult_alphas_comp}.

Because nuclear emulsion is composed of Ag, Br, O, N, and C, a realistic fragmentation model requires combining results from separate calculations for heavy (Ag, Br) and light (C, N, O) target nuclei. In order to simplify the calculations, $^{109}$Ag was used as a representative heavy target nucleus in the emulsion, and $^{12}$C was taken to represent the light nuclei. The corresponding $\alpha$-particle multiplicity distributions (Fig.~\ref{fig:mult_alphas_comp}) were then constructed by summing the calculations for $^{109}$Ag and $^{12}$C targets, weighting them by the hadronic interaction cross sections $\sigma_{\mathrm{NeAg}} = 2.57$~b and $\sigma_{\mathrm{NeC}} = 1.13$~b, and applying a correction for the emulsion atom-concentration ratio $n_{\rm light}/n_{\rm heavy} \approx 1.5$.

The calculation employing clustered Bi-pyramid density profiles for the colliding $^{22}$Ne nuclei yields fewer events without $\alpha$ particles, $N_\alpha=0$, and a higher number of events with $N_\alpha=1$ and 2. While, within experimental uncertainties, the largest fractions of events with $N_\alpha=0$ and 1 predicted by all four model variants are consistent with the data, the model does not reproduce the observed contributions of events with $N_\alpha=2$, 3, and 4. This shortcoming is consistent with the underestimation of the calculated $Z=2$ yields relative to the measurements of Ref.~\cite{Abd-Allah2001}, as seen in Fig.~\ref{fig:charge_0.9}. In the experiment~\cite{Abd-Allah2001}, multi-$\alpha$ events with $N_\alpha\ge 2$ account for roughly 20\% of all neon fragmentation events, whereas the model predicts only about 5\%. Nevertheless, one may infer that the AAMCC-MST model still reproduces the majority of $^{22}$Ne fragmentation events with $\alpha$-production. One can also expect that the presence of two extra valence neutrons in $^{22}$Ne with respect to $^{20}$Ne leads only to minor differences between the charge distributions of spectator fragments in fragmentation of $^{22}$Ne and $^{20}$Ne. This is because these two neutrons are attached randomly to produced spectator fragments and do not change their charges.

It should be emphasized that the AAMCC-MST validation results shown in Figs.~\ref{fig:charge_0.9} and \ref{fig:mult_alphas_comp} were obtained for neon fragmentation on heavy target nuclei. Because experimental data on the fragmentation of neon beams on neon targets, which would be directly relevant for the present work, are not available, measurements of neon fragmentation on light target nuclei~\cite{Zhang2012} can be used instead. Fragment production cross sections, $\sigma(Z)$ for $Z=5$–$9$, were reported in Ref.~\cite{Zhang2012} for 388$A$~MeV $^{20}$Ne fragmenting on polyethylene, carbon, and aluminum targets. Additionally, $\sigma(Z)$ for $Z=5$–$9$ were measured in Ref.~\cite{Zeitlin2011} for $^{20}$Ne beams of 290$A$, 400$A$, and 600$A$~MeV on elemental C, Al, Cu, Sn, and Pb targets.

\begin{figure}[htb!]
\centering
\includegraphics[width=0.85\textwidth]{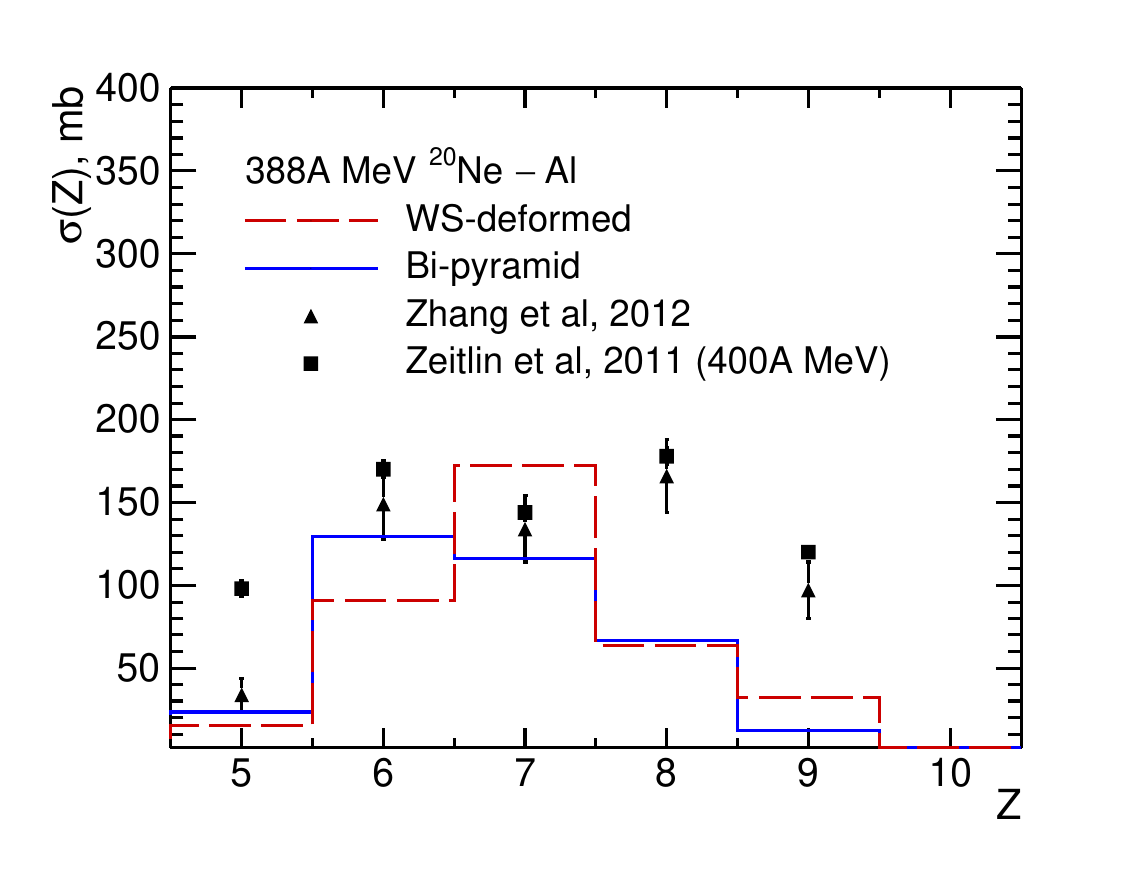}
\caption{Calculated cross sections to produce a fragment of a given charge $Z$ in fragmentation of 388$A$~MeV $^{20}$Ne on Al. AAMCC-MST results obtained with the WS-deformed and Bi-pyramid nuclear density distributions in $^{20}$Ne are presented by histograms. Experimental data for 388$A$~MeV~\cite{Zhang2012} and 400$A$~MeV~\cite{Zeitlin2011} $^{20}$Ne projectiles are presented, respectively, by triangles and squares.}
\label{fig:NeAl}
\end{figure}

In Fig.~\ref{fig:NeAl} the fragment production cross sections, $\sigma(Z)$, calculated for the fragmentation of 388$A$~MeV $^{20}$Ne on Al are presented. The measured $\sigma(Z)$ values for 388$A$~MeV~\cite{Zhang2012} and 400$A$~MeV~\cite{Zeitlin2011} $^{20}$Ne beams incident on Al are also shown in Fig.~\ref{fig:NeAl} for comparison. These two independent sets of experimental data, obtained at nearly the same beam energies, are consistent with each other within the experimental uncertainties, with the exception of the $Z=5$ cross sections, which differ significantly. The calculated cross sections for producing fragments with $Z=5$, 6, and 7 using the Bi-pyramid nuclear density distribution for $^{20}$Ne are in good agreement with the measurements of Ref.~\cite{Zhang2012}. In contrast, for $Z=8$ and 9, neither of the employed nuclear density parametrizations succeeds in reproducing the experimental results.

Overall, one may conclude that the AAMCC-MST model generally provides a satisfactory description of $^{20,22}$Ne fragmentation data~\cite{Abd-Allah2001,Neagu2013,Zhang2012,Zeitlin2011} on both light and heavy targets for all four considered calculation options. However, the model underpredicts the yields of oxygen and fluorine fragments on Al targets, as well as the rates of multi-$\alpha$ events with $N_\alpha \ge 2$ on heavy target nuclei. A more accurate description is obtained when adopting the clustered Bi-pyramid nuclear density representation for the neon projectiles and including pre-equilibrium MST clustering.

\section{Spectator fragments from \texorpdfstring{$^{20}$Ne--$^{20}$Ne}{20Ne--20Ne}  collisions at the LHC}

 After validating the AAMCC-MST model using the data on neon fragmentation on fixed targets at low collision energies, we can proceed to calculate the production of nuclear fragments in $^{20}$Ne--$^{20}$Ne collisions at $\sqrt{s_{\rm NN}} = 5.36$~TeV at the LHC. In what follows, unless explicitly stated otherwise (for instance, when discussing spectator nucleon production), the results are obtained with the pre-equilibrium MST clustering included. Based on the model validation presented above in Sec.~\ref{validation}, calculations employing the Bi-pyramid nuclear density distribution for $^{20}$Ne are preferred. However, to quantify the associated theoretical uncertainties, we also present results obtained with the Woods–Saxon spherical (WS-spherical) and Woods-Saxon deformed (WS-deformed) nuclear density parametrizations for $^{20}$Ne introduced in Sec.~\ref{initial}. 
 
\subsection{Production of \texorpdfstring{$2Z=A$}{2Z=A} fragments at the LHC}\label{2za}

This work is mainly devoted to the production of spectator fragments with $2Z = A$, namely $^2$H, $^4$He, $^6$Li, $^{10}$B, $^{12}$C, $^{14}$N, $^{16}$O, and $^{18}$F, formed in $^{20}$Ne--$^{20}$Ne collisions at the LHC. Because $^{8}$Be is extremely short-lived and rapidly decays into two $\alpha$ particles, it is excluded from the set of prefragment decay products. Unlike the light nuclei and anti-nuclei produced in nucleus--nucleus collisions at the LHC within a hot and dense fireball (considered, for example, in Ref.~\cite{Adam2015}), the spectator fragments listed above may continue to circulate in the LHC along with the $^{20}$Ne beam nuclei. Over time, these fragments can collide with beam nuclei and thus contaminate the $^{20}$Ne--$^{20}$Ne data. Such interactions are possible  since the spectator fragments are expected to retain longitudinal momenta close to those of the original beam nuclei while having very small transverse momenta.  

\begin{table}[tb!]
\caption{Cross sections of production (barns) of spectator fragments with $2Z=A$ in $^{20}$Ne--$^{20}$Ne collisions at $\sqrt{s_{\rm NN}} = 5.36$~TeV calculated with AAMCC-MST with three nuclear density parameterizations. The average values of these cross sections obtained for each kind of fragment are given in the last row with their standard deviations considered as systematic uncertainties of the calculations. The total hadronic reaction cross sections obtained with each option are given in the last column.}
\label{tab:cs_2z_eq_a}
\begin{tabular}{@{}lccccccccc@{}}
\toprule
& $^2$H & $^4$He & $^6$Li & $^{10}$B & $^{12}$C & $^{14}$N & $^{16}$O & $^{18}$F & total\\
\midrule
WS-  & 0.273 & 0.732 & 0.037 & 0.039 & 0.100 & 0.149 & 0.096 & 0.017 & 1.716 \\
spherical & & & & & & & & & \\
WS-           & 0.312 & 0.853 & 0.047 & 0.039 & 0.100 & 0.142 & 0.086 & 0.017 & 1.809 \\
deformed   & & & & & & & & & \\
Bi-             & 0.402 & 1.215 & 0.058 & 0.047 & 0.092 & 0.136 & 0.058 & 0.028 & 1.621 \\
pyramid  & & & & & & & & & \\
\midrule
average    & 0.329 & 0.933 & 0.047 & 0.042 & 0.097 & 0.142 & 0.080 & 0.021 & 1.715 \\
           & $\pm 0.066$ & $\pm 0.251$ & $\pm 0.010$ & $\pm 0.005$ & $\pm 0.005$ & $\pm 0.007$ & $\pm 0.020$ & $\pm 0.006$ & $\pm 0.09$ \\
\bottomrule
\end{tabular}
\end{table}

\begin{figure}[tb!]
    \centering
    \includegraphics[width=0.85\linewidth]{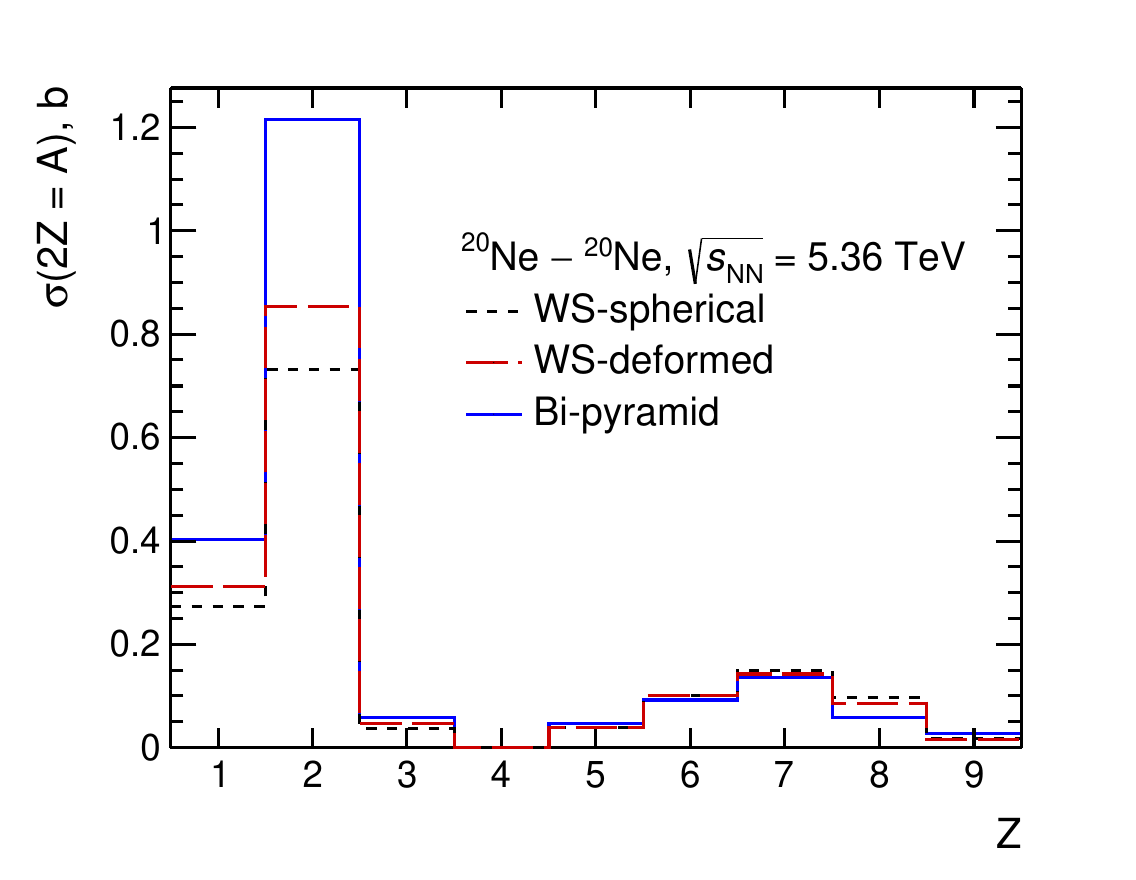}
    \caption{Calculated cross sections to produce $2Z=A$ fragments with a given charge $Z$ in  $^{20}$Ne--$^{20}$Ne collisions at $\sqrt{s_{\rm NN}}=5.36$~TeV. }
    \label{fig:sig_2z_eq_a}
\end{figure}

The cross sections $\sigma(2Z=A)$ for the production of spectator fragments with $2Z=A$ in $^{20}$Ne--$^{20}$Ne collisions at $\sqrt{s_{\rm NN}} = 5.36$~TeV were calculated using three nuclear density parameterizations: the WS-spherical, WS-deformed, and Bi-pyramid. The cross sections to produce a given kind of $2Z=A$ spectator fragment at a given side of the interaction point   were calculated per each of the colliding nuclei, and they are listed in Table~\ref{tab:cs_2z_eq_a} and shown in Fig.~\ref{fig:sig_2z_eq_a}. One can see that $^4$He fragments are the most copiously produced in $^{20}$Ne fragmentation, and their yield is maximal when the clustered Bi-pyramid density profile of $^{20}$Ne is used. Deuterons constitute the second most abundant group of $2Z=A$ fragments. Here again, the Bi-pyramid parameterization of the $^{20}$Ne density leads to the highest deuteron production cross section compared to both the WS-spherical and WS-deformed choices. $^{14}$N ranks third among the most frequently produced spectator fragments. The remaining nuclei, $^6$Li, $^{10}$B, $^{12}$C, and $^{16}$O, are generated with cross sections roughly two to three times smaller than that of $^{14}$N. Finally, within AAMCC-MST, $^{18}$F appears as the rarest $2Z=A$ fragment, formed after the removal of one proton and one neutron from $^{20}$Ne. These nucleons are either directly abraded from $^{20}$Ne or emitted via evaporation at a later stage of the reaction. This process predominantly occurs in very peripheral $^{20}$Ne--$^{20}$Ne collisions, which are characterized by relatively small cross sections. 

As shown in Table~\ref{tab:cs_2z_eq_a} and Fig.~\ref{fig:sig_2z_eq_a}, the values of $\sigma(2Z=A)$ exhibit a clear dependence on the particular choice of the nuclear density distribution in $^{20}$Ne employed in the modeling. One may therefore use the mean of the three calculations and interpret its standard deviation as a systematic modeling uncertainty, as reported in the last row of Table~\ref{tab:cs_2z_eq_a}.

In summary, the total hadronic reaction cross sections calculated for $^{20}$Ne--$^{20}$Ne collisions with the WS-spherical, WS-deformed, and Bi-pyramid  representations of nuclear density are 1.716, 1.809 and 1.621 barns, respectively, see Table~\ref{tab:cs_2z_eq_a}. At the same time, the calculated cross sections to have any of the considered $2Z=A$ fragments are 1.44, 1.60, and 2.04 barns, respectively. This means that, on average, about $84\%$ and $88\%$ of collisions modeled with WS-spherical and WS-deformed nuclear density profiles, respectively, lead to a fragment which can potentially contaminate $^{20}$Ne--$^{20}$Ne data, while using the Bi-pyramid density profile results in 1.26 of such fragments per event on average, with the most important contribution from the production of $^4$He. As seen, the production of $2Z=A$ fragments significantly increases when the Bi-pyramid density distribution with clusterization in $^{20}$Ne is employed in modeling.

\subsection{Transverse momentum distributions of \texorpdfstring{$2Z=A$}{2Z=A} fragments}\label{momenta_t}

\begin{figure*}[tb!]
    \centering
    \begin{minipage}{0.495\textwidth}
    \includegraphics[width=1.0\linewidth]{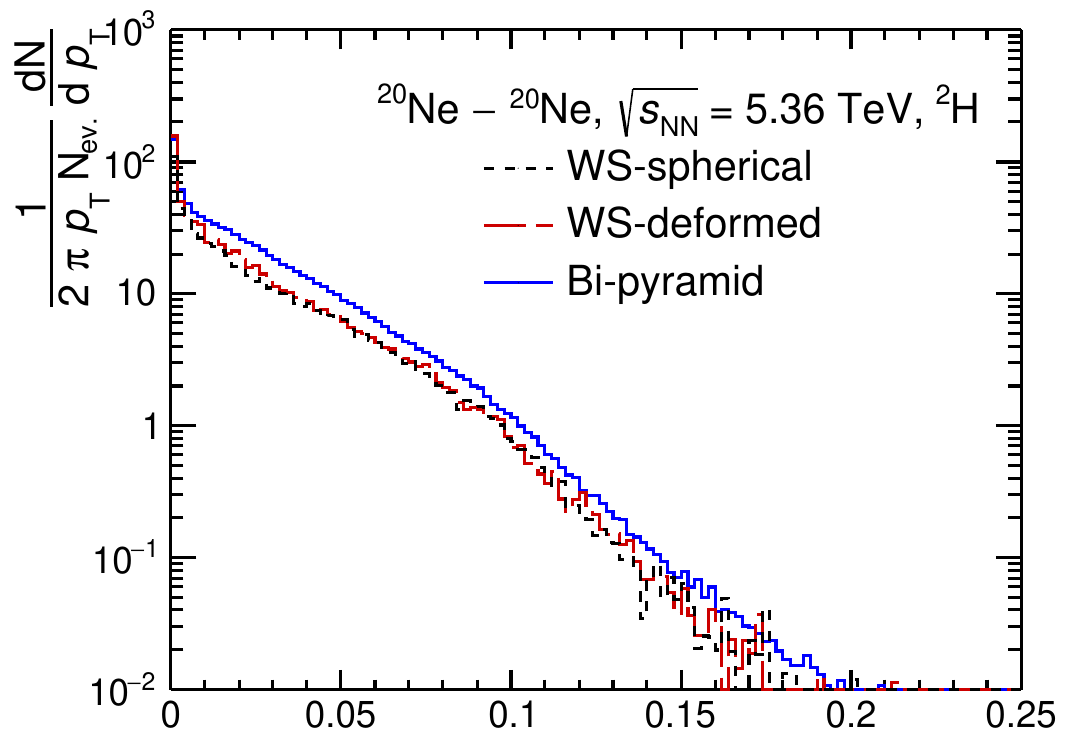}
    \end{minipage}
    \begin{minipage}{0.495\textwidth}
    \includegraphics[width=0.925\linewidth]{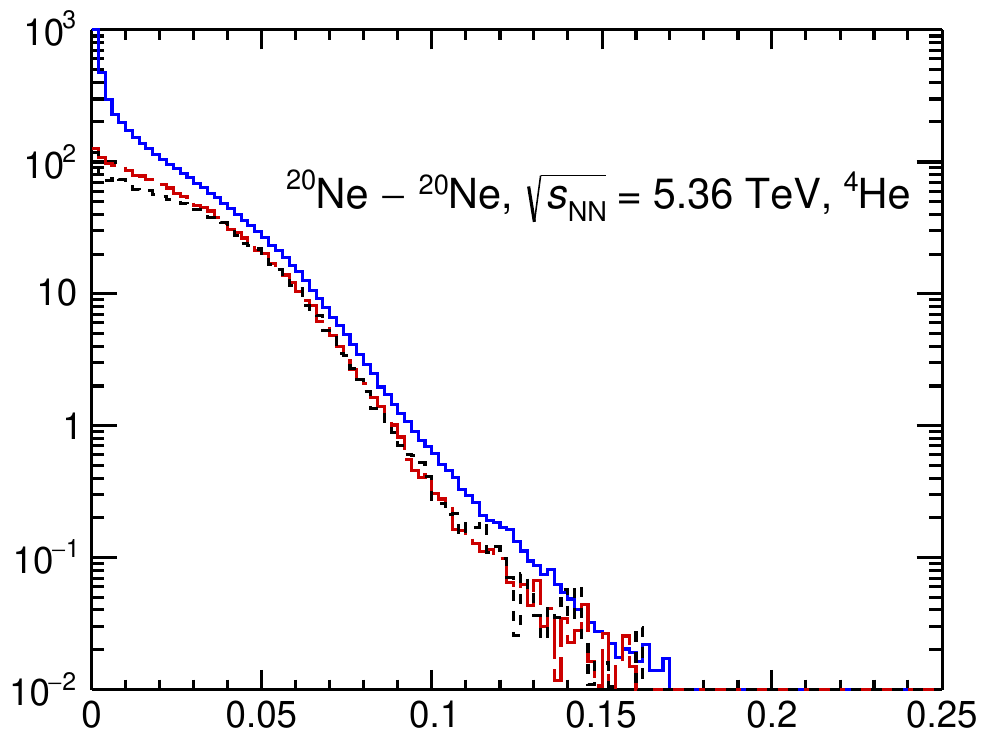}
    \end{minipage} 
    
    \begin{minipage}{0.495\textwidth}
    \includegraphics[width=1.0\linewidth]{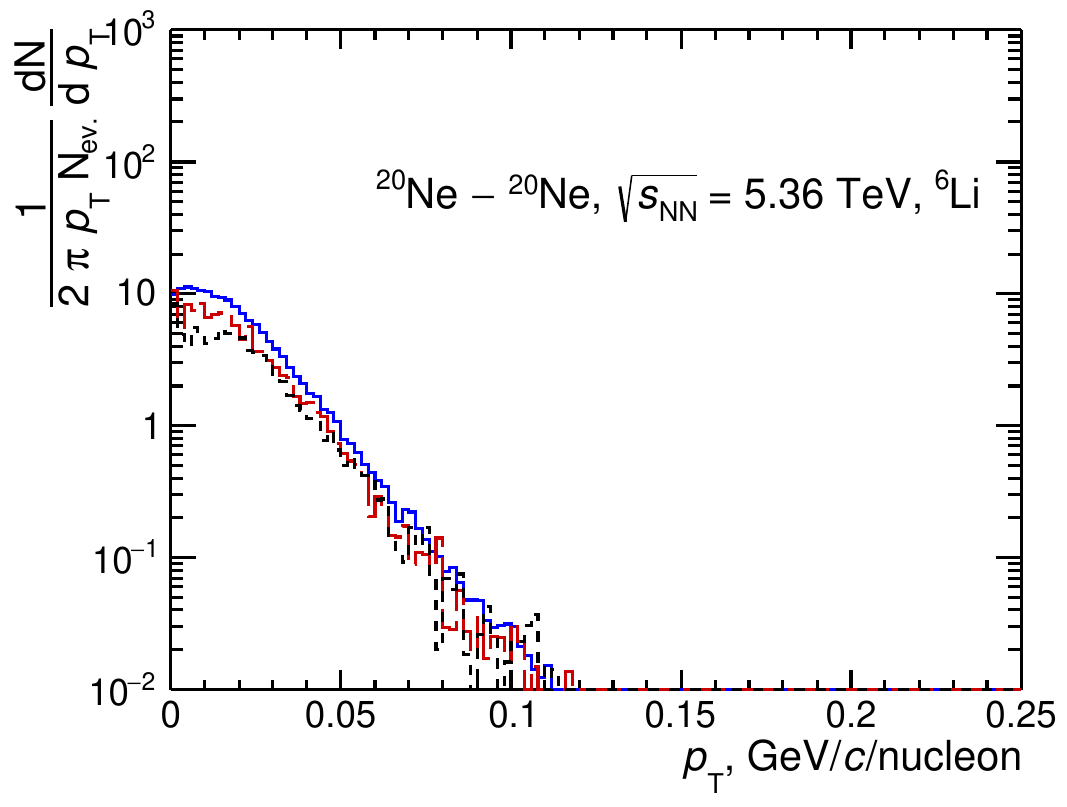}
    \end{minipage}
    \begin{minipage}{0.495\textwidth}
    \includegraphics[width=0.925\linewidth]{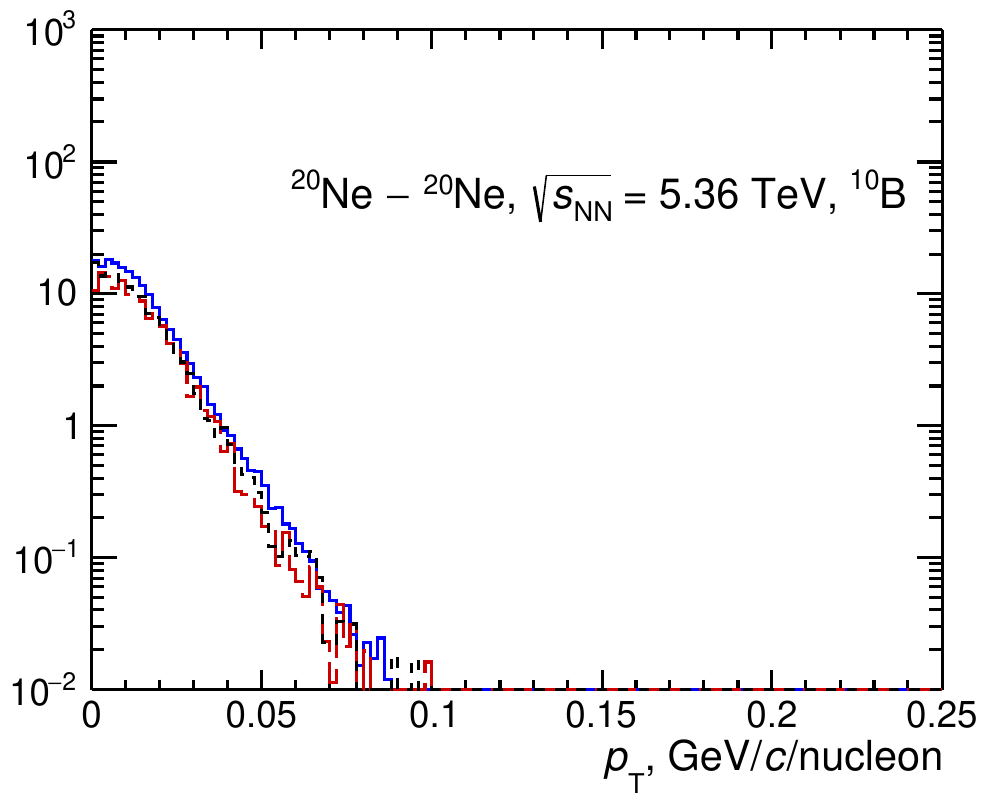}
    \end{minipage}
    \caption{Transverse momentum distributions of $2Z=A$ spectator fragments, $^2$H, $^4$He, $^6$Li, and $^{10}$B, produced in  $^{20}$Ne--$^{20}$Ne collisions at $\sqrt{s_{\mathrm{NN}}}=5.36$~TeV. The distributions were calculated per fragmenting nucleus per collision event with the AAMCC-MST model using the WS-spherical, WS-deformed, and clustered Bi-pyramid distributions of nuclear density in $^{20}$Ne.}
    \label{fig:pt_light}
\end{figure*}

\begin{figure*}[htb!]
    \centering
    \begin{minipage}{0.495\textwidth}
    \includegraphics[width=1.0\linewidth]{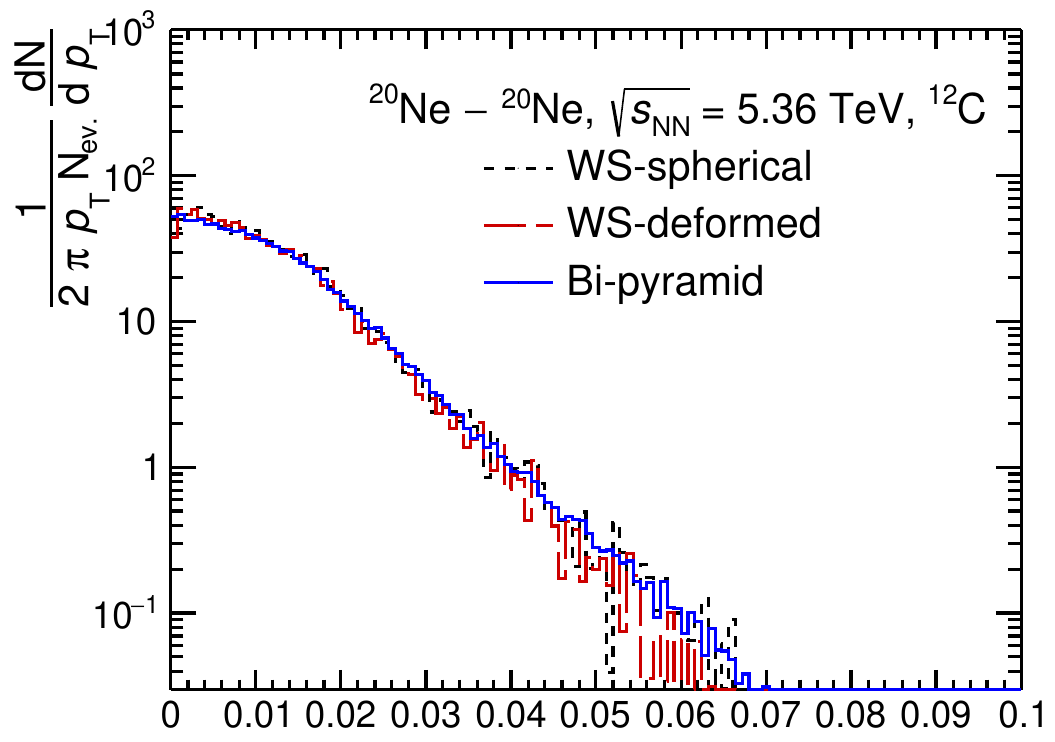}
    \end{minipage}
    \begin{minipage}{0.495\textwidth}
    \includegraphics[width=0.925\linewidth]{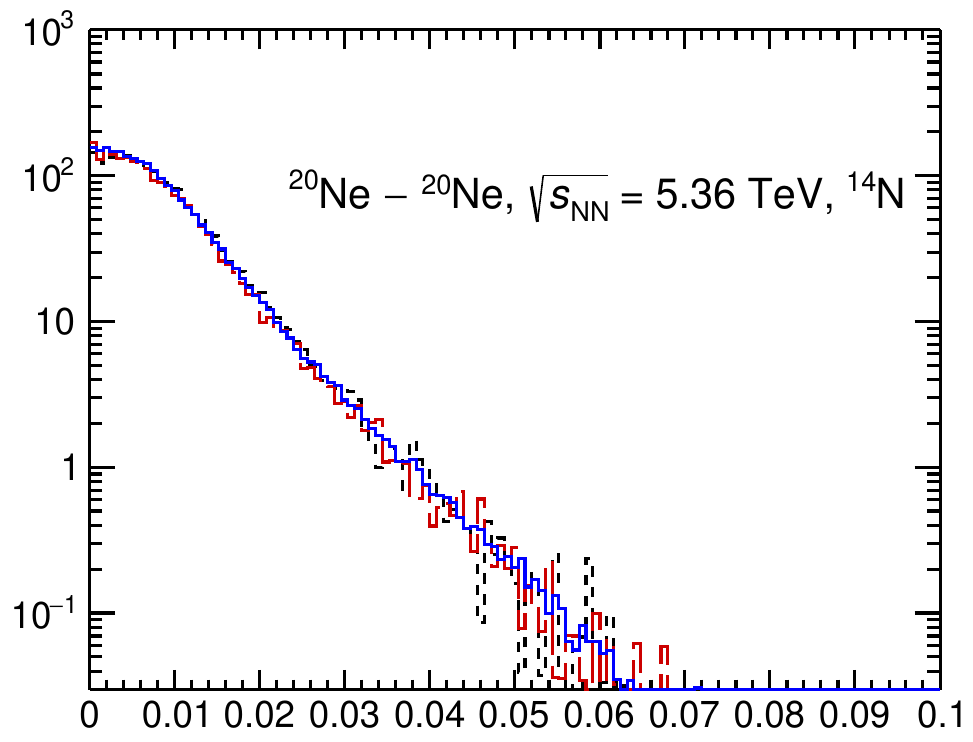}
    \end{minipage}
   
    \begin{minipage}{0.495\textwidth}
    \includegraphics[width=1.0\linewidth]{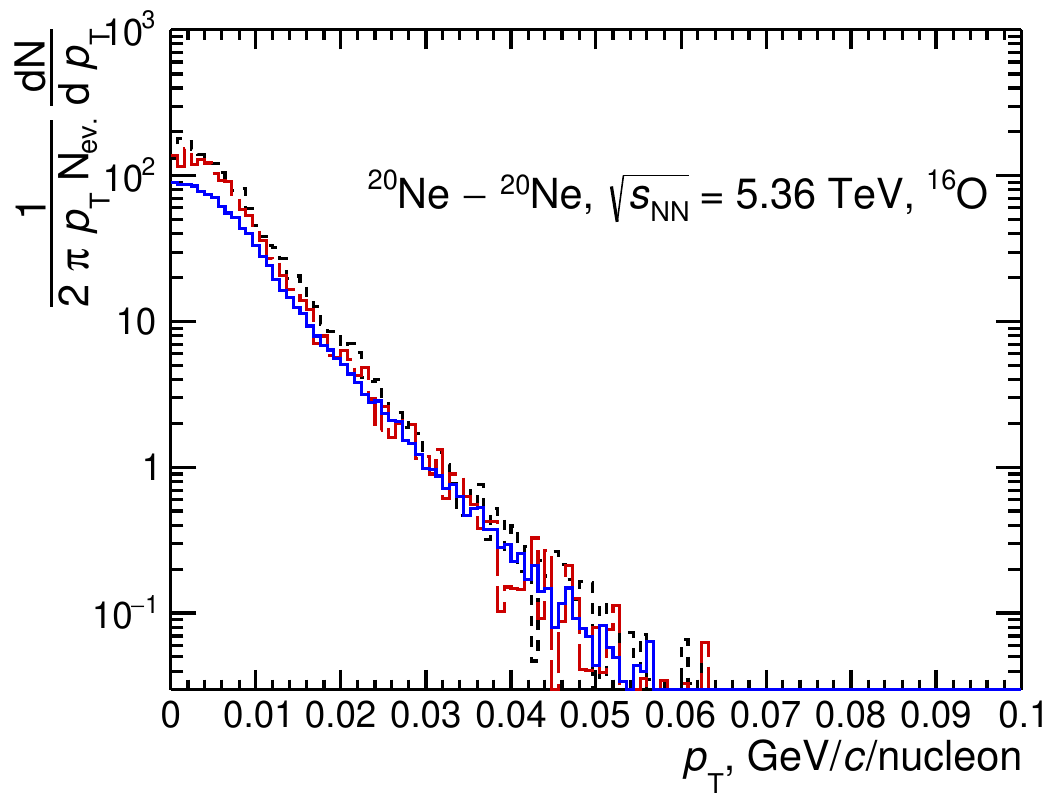}
    \end{minipage}
    \begin{minipage}{0.495\textwidth}
    \includegraphics[width=0.925\linewidth]{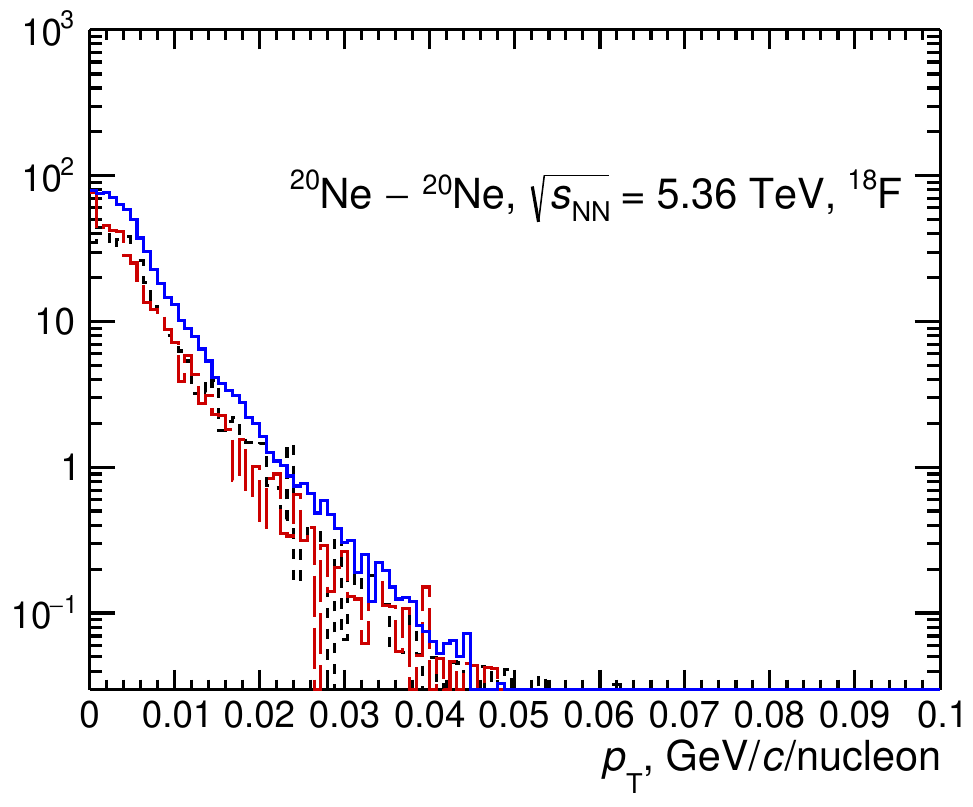}
    \end{minipage}
    \caption{Same as in Fig.~\ref{fig:pt_light}, but for $^{12}$C, $^{14}$N, $^{16}$O, and $^{18}$F spectator fragments.}
    \label{fig:pt_heavy}
\end{figure*}

After calculating the total yields of $2Z=A$ spectator fragments produced in $^{20}$Ne--$^{20}$Ne collisions, one can analyze their transverse momentum distributions. These distributions, $\tfrac{1}{2\pi p_{\mathrm T} N_{\mathrm ev.}}\tfrac{\mathrm dN}{{\mathrm d}p_{\mathrm T}}$, are shown in Figs.~\ref{fig:pt_light} and~\ref{fig:pt_heavy} as functions of the transverse momentum per nucleon of the spectator fragment, $p_{\mathrm T}$. The distributions were calculated at one side, i.e., per fragmenting $^{20}$Ne nucleus, per collision event. The $p_{\mathrm T}$ distributions for $^2$H, $^4$He, $^6$Li, and $^{10}$B in Fig.~\ref{fig:pt_light}, and for $^{12}$C, $^{14}$N, $^{16}$O, and $^{18}$F in Fig.~\ref{fig:pt_heavy}, reflect the differences between the cross sections obtained with different $^{20}$Ne density options and listed in Table~\ref{tab:cs_2z_eq_a}. The $p_{\mathrm T}$ distributions calculated for $^{10}$B, $^{12}$C, and $^{14}$N fragments are nearly identical for all three $^{20}$Ne density distribution options, whereas the $p_{\mathrm T}$ distributions for $^2$H, $^4$He, and $^{18}$F obtained with the clustered Bi-pyramid distribution in $^{20}$Ne lie above those predicted with the other two density options.

As seen in Fig.~\ref{fig:pt_light}, the simulations that employ the clustered Bi-pyramid nuclear density distribution for $^{20}$Ne indicate an enhanced yield of $^4$He fragments at low $p_{\mathrm T}$. The $p_{\mathrm T}$ distributions obtained with the WS-spherical and WS-deformed density profiles for $^{20}$Ne are almost the same across all spectator fragment species, including $^2$H, $^4$He, and $^{16}$O, $^{18}$F fragments. Nevertheless, the  $p_{\mathrm T}$ distributions calculated with the WS parameterizations differ not only in magnitude, but also in shape from those obtained with the Bi-pyramid density, with the exception of $^{10}$B, $^{12}$C, and $^{14}$N, see Fig.~\ref{fig:pt_light} and~\ref{fig:pt_heavy}.

It is also evident that the $p_{\mathrm T}$ distributions become narrower with increasing fragment mass. Although some $^2$H and $^4$He fragments are produced with transverse momenta up to 150--200~MeV/$c$/nucleon, the transverse momenta of $^{16}$O and $^{18}$F fragments remain below 40--50~MeV/$c$/nucleon, see Figs.~\ref{fig:pt_light} and \ref{fig:pt_heavy}. Thus, if one assumes that the longitudinal momenta $p_z$ of all produced spectator fragments calculated per fragment nucleon are close to the nominal beam momentum, it follows that all $^2$H and $^4$He fragments are confined within an emission cone of 56--75~$\mu$rad, while all $^{16}$O and $^{18}$F fragments are emitted into a much narrower cone of 15--19~$\mu$rad.

These angular spreads can be compared with the characteristic RMS angular divergence of the initial $^{20}$Ne beams at the interaction points of the LHC experiments. During the 2025 oxygen--oxygen and neon--neon runs, the beam optics configuration employed a beta function of $\beta^* = 0.5$~m at the interaction points of ALICE (IP2), ATLAS (IP1), and CMS (IP5)~\cite{Bruce2026}. Assuming a typical normalized emittance in the neon run of $\epsilon_n \approx 3$~$\mu$m$\cdot$rad, to be similar to the value previously measured for $^{208}$Pb beams~\cite{Sedlacek2025}, the corresponding RMS angular divergence at the IPs is $\sigma_{x\prime} \approx 46$~$\mu$rad. Consequently, the majority of $^2$H and $^4$He fragments, as well as all $^{16}$O and $^{18}$F fragments, are expected to remain within the beam envelope, at least in the vicinity of IP2, IP1, and IP5. Again, this estimation is based on the simple assumption that the longitudinal momenta of fragments $p_z$, calculated per fragment nucleon, are close to the nominal beam momentum. This assumption is tested against calculated $p_z$ distributions of $2Z = A$ spectator fragments below in Sec.~\ref{momenta_l}.

\subsection{Longitudinal momentum distributions of \texorpdfstring{$2Z=A$}{2Z=A} fragments}\label{momenta_l}

\begin{figure*}[tb!]
    \centering
    \begin{minipage}{0.495\textwidth}
    \includegraphics[width=1.0\linewidth]{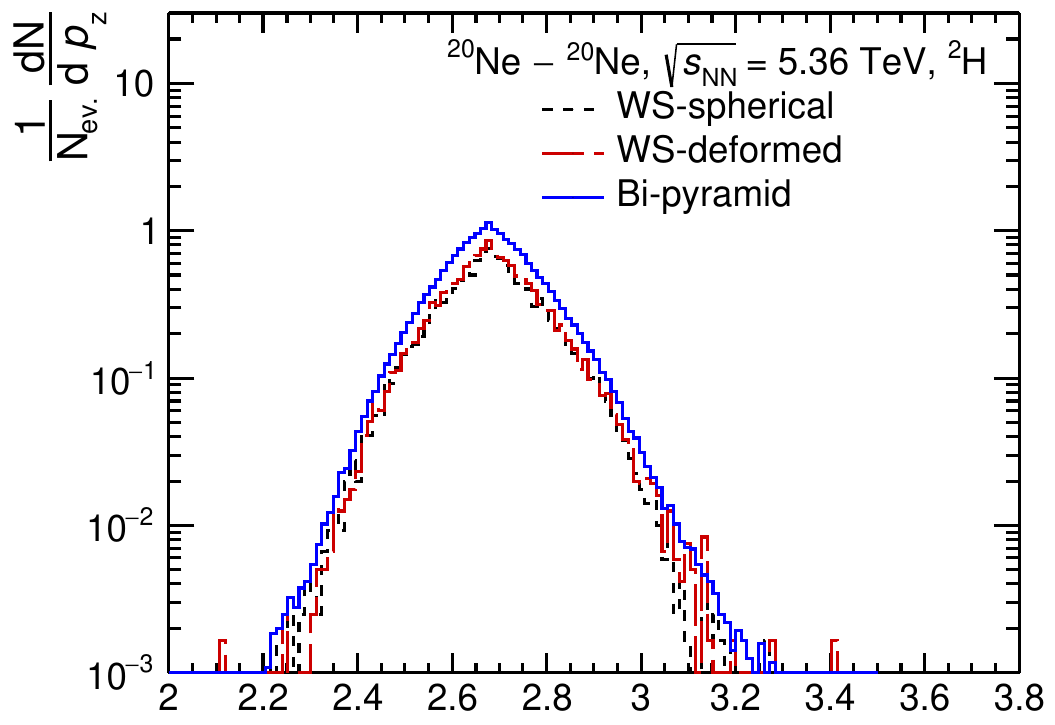}
    \end{minipage}
    \begin{minipage}{0.495\textwidth}
    \includegraphics[width=0.925\linewidth]{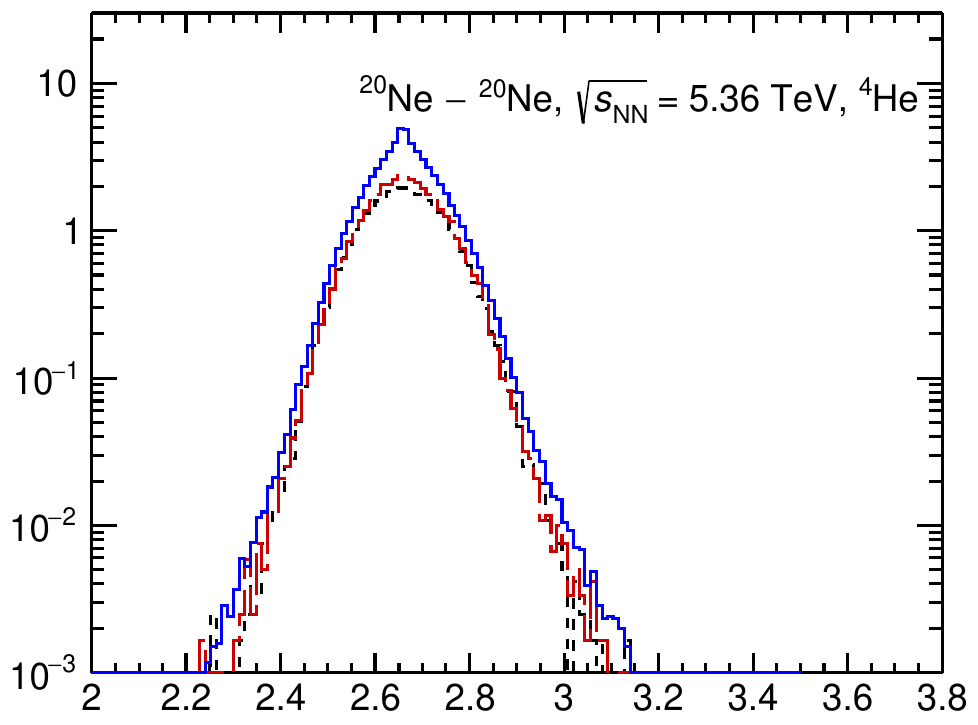}
    \end{minipage}
   
    \begin{minipage}{0.495\textwidth}
    \includegraphics[width=1.0\linewidth]{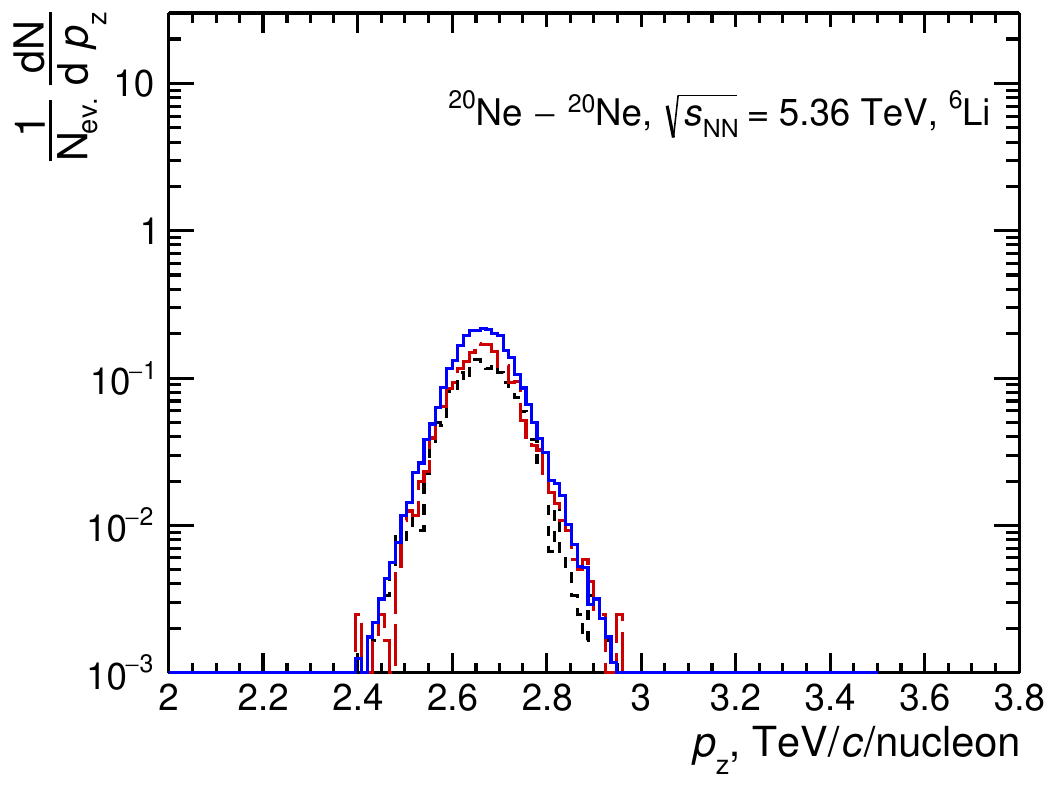}
    \end{minipage}
    \begin{minipage}{0.495\textwidth}
    \includegraphics[width=0.925\linewidth]{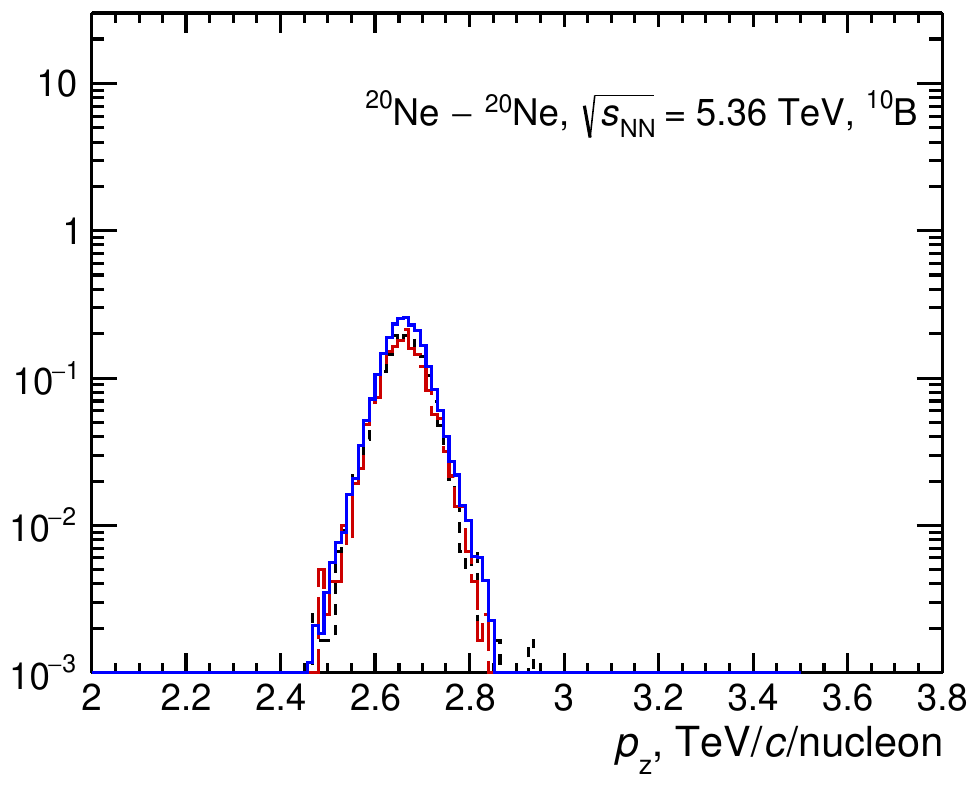}
    \end{minipage}
    \caption{Longitudinal momentum distributions of $2Z=A$ spectator fragments: $^2$H, $^4$He, $^6$Li, and $^{10}$B,  produced in  $^{20}$Ne--$^{20}$Ne collisions at $\sqrt{s_{\mathrm{NN}}}=5.36$~TeV. The distributions were calculated per fragmenting nucleus and per collision event with the AAMCC-MST model using the WS-spherical, WS-deformed, and clustered Bi-pyramid distributions of nuclear density in $^{20}$Ne.}
    \label{fig:pz_light}
\end{figure*}

\begin{figure*}[tb!]
    \centering
    \begin{minipage}{0.495\textwidth}
    \includegraphics[width=1.0\linewidth]{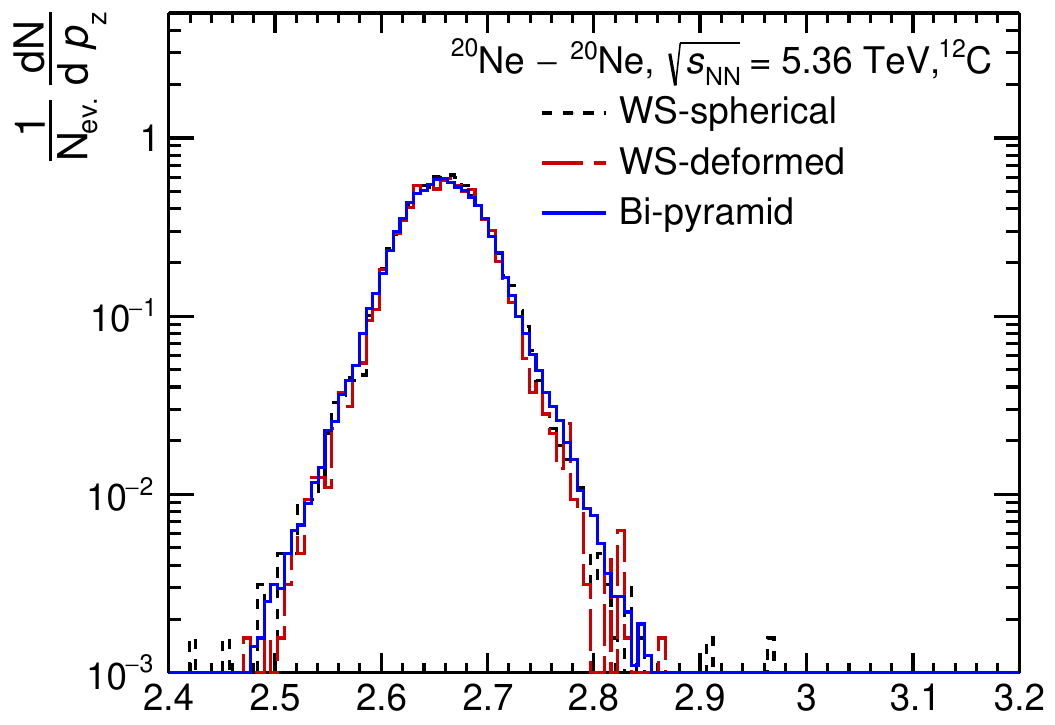}
    \end{minipage}
    \begin{minipage}{0.495\textwidth}
    \includegraphics[width=0.925\linewidth]{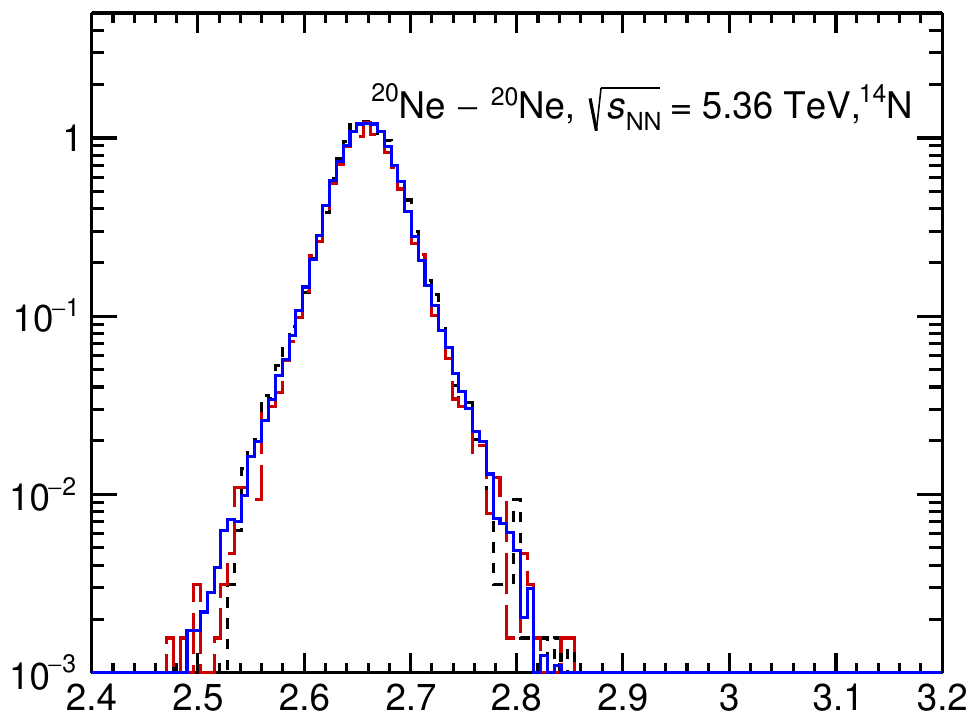}
    \end{minipage}
  
    \begin{minipage}{0.495\textwidth}
    \includegraphics[width=1.0\linewidth]{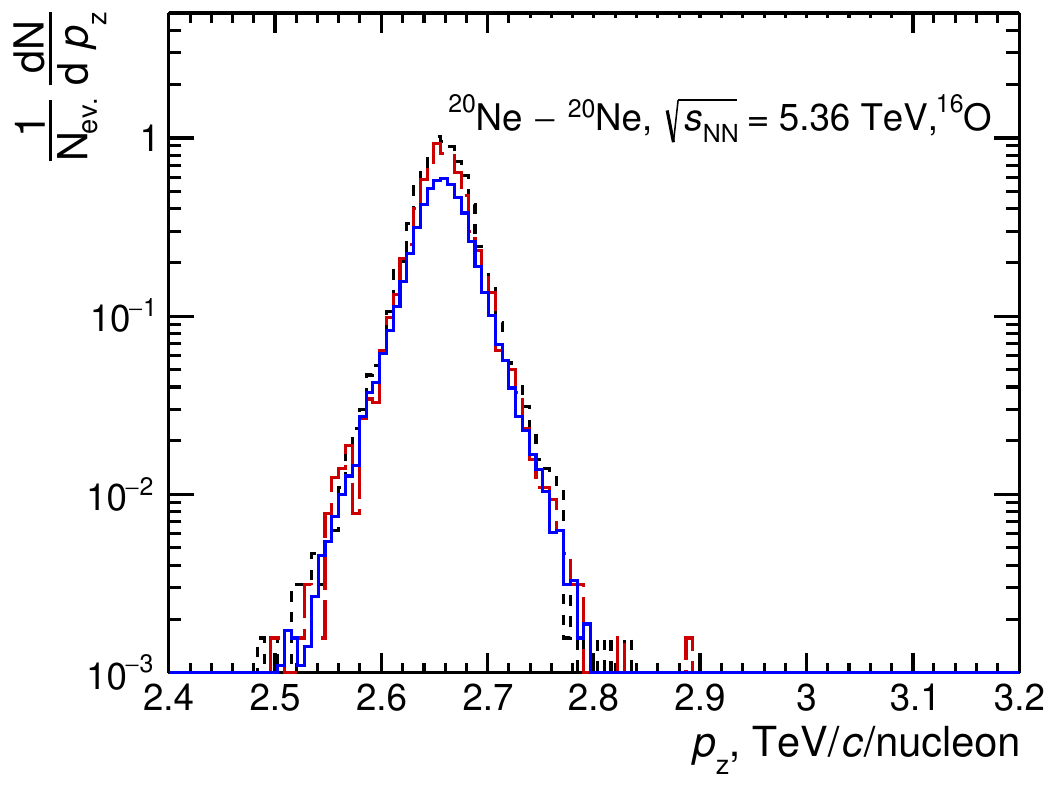}
    \end{minipage}
    \begin{minipage}{0.495\textwidth}
    \includegraphics[width=0.925\linewidth]{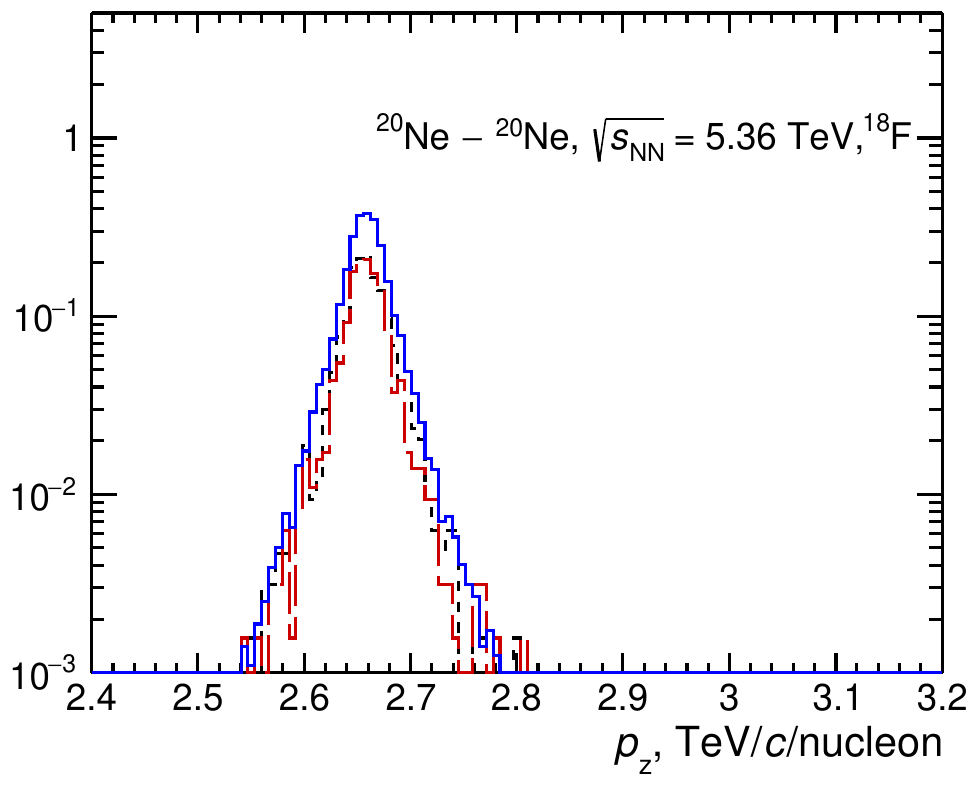}
    \end{minipage}
    \caption{Same as in Fig.~\ref{fig:pz_light}, but for $^{12}$C, $^{14}$N, $^{16}$O, and $^{18}$F spectator fragments.}
    \label{fig:pz_heavy}
\end{figure*}

The calculated with AAMCC-MST longitudinal momentum, $p_z$, distributions of $2Z=A$ spectator fragments, $\tfrac{1}{N_{\mathrm ev.}}\tfrac{\mathrm dN}{{\mathrm d}p_z}$, produced in $^{20}$Ne--$^{20}$Ne collisions at $\sqrt{s_{\mathrm{NN}}}=5.36$~TeV are presented in Figs.~\ref{fig:pz_light} and~\ref{fig:pz_heavy}. The distributions were calculated at one side only, i.e., per fragmenting $^{20}$Ne nucleus and per collision event. These results were obtained using the WS-spherical, WS-deformed, and clustered Bi-pyramid distributions of nuclear density in $^{20}$Ne. 

The relationships among the $p_z$ distributions calculated for different spectator species using the WS-spherical, WS-deformed, and clustered Bi-pyramid density distributions in $^{20}$Ne are similar to those found for the $p_{\mathrm T}$ distributions, see Sec.~\ref{momenta_t}. For all spectator fragment species, the $p_z$ distributions obtained with the WS-spherical and WS-deformed density profiles for $^{20}$Ne are nearly identical. However, the $p_{\mathrm z}$ distributions calculated with the WS parameterizations differ from those based on the Bi-pyramid density not only in magnitude but also in shape, except for $^{10}$B, $^{12}$C, and $^{14}$N, see Fig.~\ref{fig:pz_light} and~\ref{fig:pz_heavy}. Consequently, in the following, only the results from calculations employing the WS-deformed and Bi-pyramid density distributions in $^{20}$Ne are presented.

\begin{figure}
    \centering
    \includegraphics[width=0.85\linewidth]{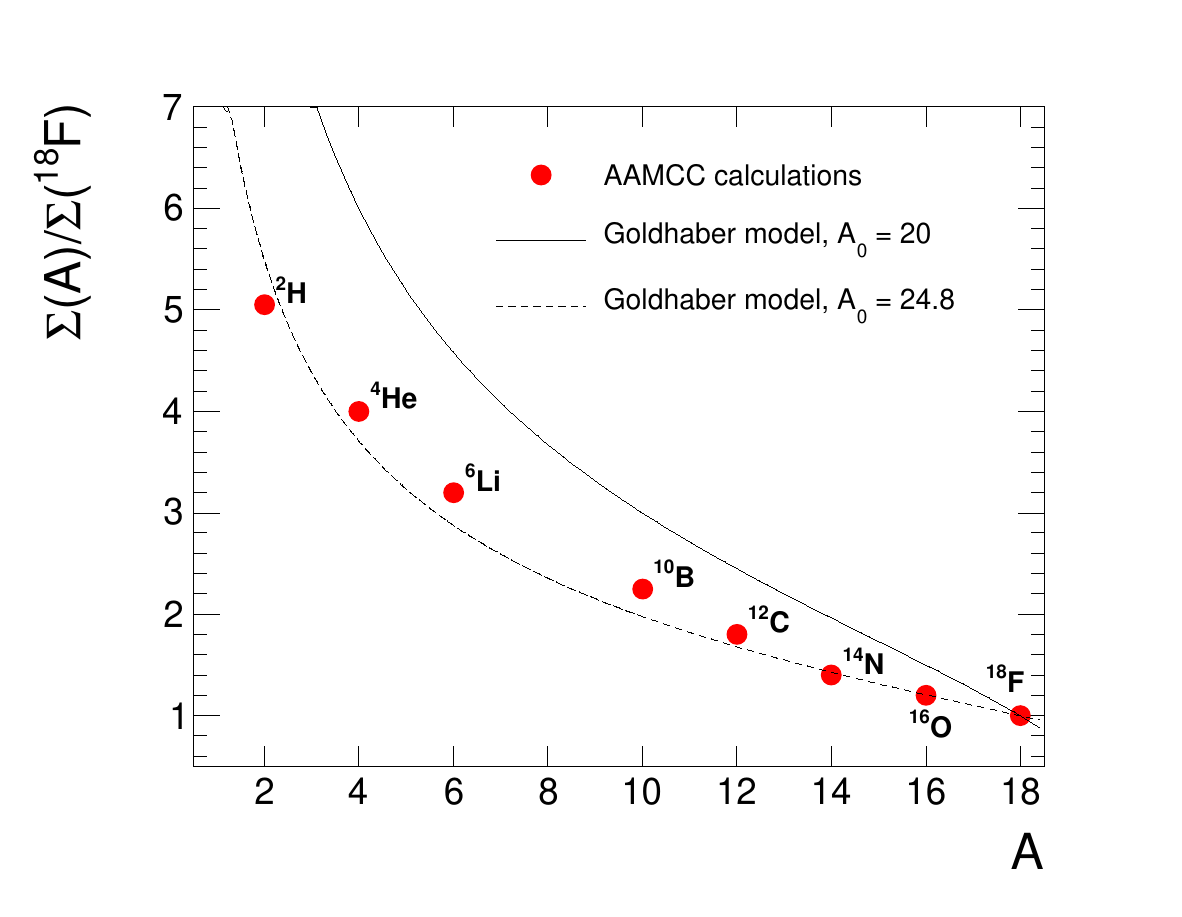}
    \caption{RMS deviations of longitudinal momenta $p_z$ per nucleon of spectator $2Z=A$ fragments from the nominal beam momentum $p_b=2.68$~TeV/$c$ calculated relative to RMS for $^{18}$F,  $\Sigma/\Sigma(^{18}\mathrm{F})$, as functions of fragment mass number $A$ for $^{20}$Ne--$^{20}$Ne collisions at $\sqrt{s_{\rm NN}}=5.36$~TeV. Points present results of the Gaussian fit to $p_z$ distributions calculated with AAMCC-MST using Bi-pyramid density distribution in $^{20}$Ne. The fitting errors fall within the size of the markers. An estimation based on the Goldhaber model with $A_0 = 20$ is presented by a solid curve, see text for details. A similar estimation, but with the effective mass $A_0 = 24.8$ to fit AAMCC-MST results is presented by a dashed curve.}
    \label{fig:sigmas}
\end{figure}

All distributions shown in Figs.~\ref{fig:pz_light} and~\ref{fig:pz_heavy} are naturally centered at the nominal beam momentum of  $p_b=2.68$~TeV/$c$/nucleon, since the spectator fragments are the remnants of $^{20}$Ne beam nuclei. In the rest frame of the fragmenting $^{20}$Ne projectile, these spectators are typically emitted with momenta of several  hundred MeV/$c$/nucleon. They can be emitted in various directions, including along the beam direction and opposite to it. After applying the Lorentz boost to the laboratory frame, these forward- and backward-emitted fragments populate the right and left sides of the distributions in Figs.~\ref{fig:pz_light} and~\ref{fig:pz_heavy}, producing a finite spread in $p_z$ around the nominal beam momentum. Owing to their symmetric shapes, the distributions can be approximated with Gaussian functions having a mean of $p_b=2.68$~TeV/$c$/nucleon and an RMS width $\Sigma$, determined individually for each of the considered $2Z=A$ fragment species. 

The RMS width values, $\Sigma$, extracted from Gaussian fits to the $p_z$ distributions of spectator fragments produced in $^{20}$Ne--$^{20}$Ne collisions are summarized in Table~\ref{tab:sig_pz}. One can note that the differences between the values of $\Sigma$ obtained with different density options are rather small. The RMS widths of the $p_z$ distributions calculated for the considered $2Z=A$ fragments, normalized to the RMS width of the $^{18}$F, are presented in Fig.~\ref{fig:sigmas}. As seen, $\Sigma$ decreases as the fragment mass increases. Specifically, the RMS width of the $p_z$ distribution for $^{18}$F is about five times smaller than that for $^2$H.

\begin{table}[htb!]
\caption{Standard deviations $\Sigma$ (TeV/$c$/nucleon) extracted from Gaussian fits to the longitudinal momentum ($p_z$, calculated per fragment nucleon) distributions of $2Z=A$ spectator fragments produced in $^{20}$Ne--$^{20}$Ne collisions at $\sqrt{s_{\rm NN}}=5.36$~TeV. These distributions were calculated using the AAMCC-MST model, taking the nuclear densities in $^{20}$Ne to be either the WS-deformed or Bi-pyramid, and $\Sigma$ values were determined independently for each case.}\label{tab:sig_pz}
\begin{tabular}{@{}lcccccccc@{}}
\toprule
& $^2$H & $^4$He & $^6$Li & $^{10}$B & $^{12}$C & $^{14}$N & $^{16}$O & $^{18}$F\\
\midrule
WS-deformed & $0.110$   & $0.084$  & $0.066$ & $0.049$ & $0.038$ & $0.029$ &  $0.024$ &  $0.018$\\
Bi-pyramid            & $0.111$   & $0.080$  & $0.067$ & $0.048$ & $0.040$ & $0.029$ &  $0.026$ &  $0.020$\\
\botrule
\end{tabular}
\end{table}

Due to secondary decays of prefragments, as well as clusters formed after the pre-equilibrium MST-clustering of spectator matter, the widths of momentum distributions of final fragments are expected to deviate from the dependence on fragment mass number $A$ implied by Eq.~(\ref{eq:Goldhaber_sigma}). To obtain the modified dependence, Eq.~(\ref{eq:Goldhaber_sigma}) can 
be recast first for the distribution of the fragment longitudinal momentum $p_z$ calculated per nucleon, rather than for the distribution of the total fragment momentum $P_z$. By introducing mean values, $P_b$ and $p_b$, and RMS widths, $\sigma$ and $\Sigma$, for the total-momentum and per-nucleon-momentum distributions, respectively, of a fragment with mass number $A$, the Gaussian distribution can be written as:
\begin{equation}
    \exp{\Big[-\frac{(P_z-P_b)^2}{2\sigma^2}\Big]}= \exp{\Big[-\frac{A^2(p_z-p_b)^2}{2\sigma^2}\Big]}= \exp{\Big[-\frac{(p_z-p_b)^2}{2\Sigma^2}\Big]}  \, 
\end{equation}
and, therefore, following Eq.~(\ref{eq:Goldhaber_sigma}):
\begin{equation}
\Sigma^2=\frac{\sigma^2}{A^2}=\frac{\sigma_0^2A(A_0-A)}{A^2(A_0-1)}=\frac{\sigma_0^2}{A_0-1}\Big(\frac{A_0}{A}-1  \Big) \ ,    
\end{equation}
where $A_0$ is considered as an effective mass number of a decaying system. 

It is convenient to consider the RMS width of each $p_z$ distribution relative to the RMS width of the distribution for $^{18}$F and present the ratio as a function of the fragment mass number $A$:  
\begin{equation}
    \frac{\Sigma(A)}{\Sigma(^{18}\mathrm{F})}=\sqrt{\frac{A_0/A-1}{A_0/18-1}}\ .
\label{eq:sigma_rel}
\end{equation}

As a first approximation, using Eq.~(\ref{eq:Goldhaber_sigma}), one may set $A_0=20$, thereby treating all considered $2Z=A$ fragment species as fragments of $^{20}$Ne, and neglecting their decays. However, as shown in Fig.~\ref{fig:sigmas}, the curve obtained in this way systematically overpredicts the relative RMS widths for $2Z=A$ fragments with respect to those calculated with  AAMCC-MST. Since the prefragment momentum is redistributed among its decay products, the momentum distributions of the final-state fragments are expected to be narrower than the original prefragment momentum distribution. A good agreement with the AAMCC-MST results is achieved by increasing the parameter from $A_0=20$ to $A_0=24.8$, which effectively decreases the relative RMS widths (see Fig.~\ref{fig:sigmas}). As a result, the dependence of the relative RMS width of the $p_z$ distribution for a $2Z=A$ fragment on its mass number, as obtained from the full AAMCC-MST modeling, is well captured by the simple relation given by Eq.~(\ref{eq:sigma_rel}).  

As reported in~Ref.~\cite{Apollinari2015}, the LHC delivers proton and lead-ion beams with a very small RMS longitudinal-momentum (energy) spread, at the level of $\sigma_{p_z}/p_z \sim 1.13\cdot 10^{-4}$. This is provided by the advanced LHC beam optics, which effectively filters out and removes nuclei whose $p_z$ differs too strongly from the nominal beam momentum. As noted in Ref.~\cite{Nijs2025}, in recent $^{16}$O--$^{16}$O and $^{20}$Ne--$^{20}$Ne runs, the majority of secondary nuclei with $2Z=A$ failed to circulate because the optics configuration accepted only deviations up to $|\delta p_z|/p_z \sim 3.4\cdot 10^{-4}$ with respect to the nominal momentum. Such successful beam cooling provides the planned design value $3\sigma_{p_z}/p_z$ reported in Ref.~\cite{Apollinari2015}. 

As follows from Table~\ref{tab:sig_pz}, the $\Sigma_{p_z}/p_z$ values derived from the AAMCC-MST modeling are markedly larger, namely 
$414\cdot 10^{-4}$ for $^2$H and $74\cdot 10^{-4}$ for $^{18}$F. Consequently, most of the considered $2Z=A$ fragments will be removed from $^{20}$Ne beams. An accurate evaluation of the fraction of $2Z=A$ fragments that can circulate in the LHC would require a detailed simulation of the LHC beam optics, which lies outside the scope of the present work, whose aim is to predict the total  yields and momentum distributions of $2Z=A$ fragments.

\subsection{Multiplicity distribution of helium fragments}\label{he}

\begin{figure}[htb!]
    \centering
    \includegraphics[width=0.85\linewidth]{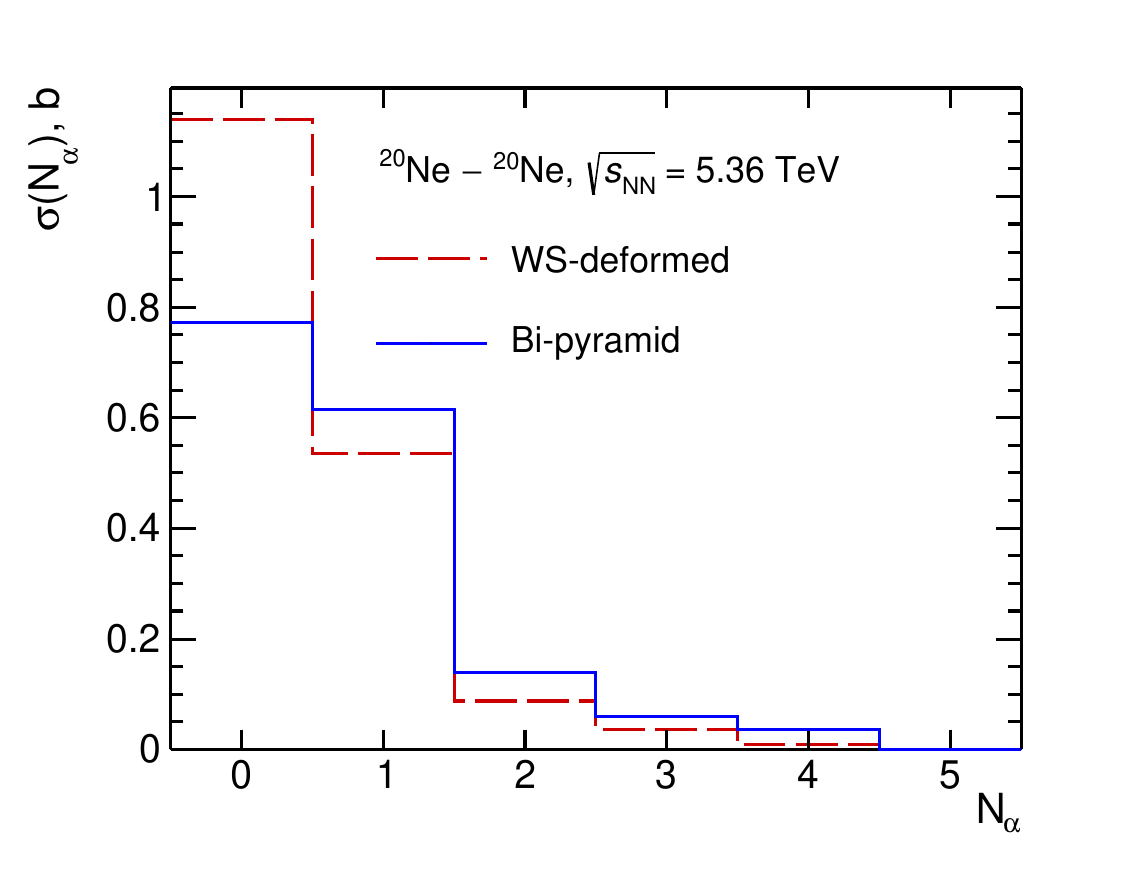}
    \caption{Cross sections of production of a given number of spectator $\alpha$ particles in $^{20}$Ne--$^{20}$Ne collisions at $\sqrt{s_{\rm NN}}=5.36$~TeV calculated with the WS-deformed and Bi-pyramid distributions of nuclear density in  $^{20}$Ne.}
    \label{fig:mult_alpha}
\end{figure}

Because a single $^{20}$Ne fragmentation event can yield multiple $^4$He ($\alpha$ particles), these fragments constitute the most abundant $2Z=A$ products. The production cross sections for events with a given multiplicity of spectator $\alpha$ particles were calculated with AAMCC-MST, assuming either the WS-deformed or  Bi-pyramid nuclear-density distribution for $^{20}$Ne, and results are shown in Fig.~\ref{fig:mult_alpha}. With the WS-deformed density, the predicted cross section for events without $\alpha$ particles ($N_\alpha=0$) is about 50\% higher than that obtained with the Bi-pyramid density. Compared with the simulations of $^{22}$Ne fragmentation on heavy target nuclei (Fig.~\ref{fig:mult_alphas_comp}), the production of spectator $\alpha$ particles is more frequent in $^{20}$Ne--$^{20}$Ne collisions when the Bi-pyramid option is used. This indicates that preformed spectator $\alpha$ particles in the Bi-pyramid configuration of $^{20}$Ne have a higher probability to survive in interactions with a lighter and smaller target nucleus than with a larger and heavy target, where they frequently split into spectator nucleons. This also demonstrates a high sensitivity of the production of spectator $\alpha$ particles in $^{20}$Ne--$^{20}$Ne collisions at the LHC to details of nuclear structure of colliding nuclei.

\section{Free spectator neutrons and protons and their detection by Zero Degree Calorimeters}\label{nucleons}

After assessing the yields of $2Z=A$ spectator fragments as possible contaminants of $^{20}$Ne beams, it is necessary to consider the yields of free spectator neutrons and protons calculated with AAMCC-MST. Detecting forward neutrons is one of the methods used by the LHC experiments, in particular, by ALICE~\cite{Abelev2013}, to determine the centrality of hadronic $^{208}$Pb--$^{208}$Pb collisions. LHC luminosity measurements in $^{208}$Pb--$^{208}$Pb runs~\cite{Acharya2024lumi} also rely in part on registering at least one neutron in the neutron ZDCs on either side of the IP. Such neutrons are produced either in hadronic $^{208}$Pb--$^{208}$Pb interactions or via electromagnetic dissociation (EMD)~\cite{Pshenichnov2011} of $^{208}$Pb in ultraperipheral collisions~\cite{Acharya2023}. The lack of forward spectator neutrons in a hadronic $^{208}$Pb--$^{208}$Pb event is very improbable, since both nuclei usually emit numerous spectator neutrons, and neutron emission occurs in $\sim 97$\% of ultraperipheral EMD events~\cite{Pshenichnov2011}. Therefore, observing a neutron on at least one side of the IP serves as a reliable trigger for both hadronic and EMD interactions of $^{208}$Pb nuclei at the LHC.

Forward neutron and proton production in $^{208}$Pb--$^{208}$Pb interactions at the LHC has been extensively investigated, including forward nucleon emission in EMD~\cite{Acharya2023,Acharya2025}. In contrast, forward neutrons and protons produced via hadronic fragmentation of $^{16}$O and $^{20}$Ne at the LHC remain much less explored. In our study of $^{16}$O fragmentation in hadronic $^{16}$O--$^{16}$O collisions at the LHC, also using AAMCC-MST~\cite{Svetlichnyi2023}, we found that a significant fraction of $^{16}$O--$^{16}$O events contain no spectator neutrons (33–43\%). As a result, the forward neutron ZDCs currently installed at the LHC, when used alone, are expected to be less effective for determining collision centrality and for luminosity monitoring during the $^{16}$O--$^{16}$O runs.

The capability of neutron and proton ZDCs to detect $^{20}$Ne--$^{20}$Ne collisions at the LHC can be assessed using the modeling results summarized in Table~\ref{tab:np_mult}. In simulations employing the WS-deformed and Bi-pyramid density profiles for $^{20}$Ne, fragmentation channels with the emission of both neutrons and protons ($\mathrm{N_n}>0,\mathrm{N_p}>0$) account for 62.4\% and 40.1\% of the total cross section, respectively. In addition, the shares of events with neutrons but without protons ($\mathrm{N_n}>0,\mathrm{N_p}=0$) are 11.1\% and 14.9\%, respectively. 

\begin{table}[htbp!]
\centering
\caption{Cross sections for fragmentation channels with emission of at least one neutron and also at least one proton, $\sigma(\mathrm{N_n}>0,\mathrm{N_p}>0)$, in $^{20}$Ne fragmentation in $^{20}$Ne--$^{20}$Ne collisions at $\sqrt{s_{\rm NN}}=5.36$~TeV calculated with the WS-deformed and Bi-pyramid nuclear density distributions for $^{20}$Ne. Other cross sections are given as well: for channels with emission of free spectator neutrons but without protons; conversely, with protons but without neutrons; and for channels in which neither free neutrons nor free protons are emitted. The corresponding channel probabilities are also listed.}
\begin{tabular}{ccccc}
\hline
& \multicolumn{2}{c}{WS-deformed} & \multicolumn{2}{c}{Bi-pyramid} \\
Channel & $\sigma$, b & Probability, \% & $\sigma$, b & Probability, \% \\
\hline
${\mathrm{N_n}}>0,\ {\mathrm{N_p}}>0$
  & 1.129 & 62.4
  & 0.650 & 40.1 \\
${\mathrm{N_n}}> 0,\ {\mathrm{N_p}}=0$
  & 0.201 & 11.1
  & 0.242 & 14.9 \\
${\mathrm{N_n}}=0,\ {\mathrm{N_p}}>0$
  & 0.345 & 19.1
  & 0.420 & 25.9 \\
${\mathrm{N_n}}=0,\ {\mathrm{N_p}}=0$
  & 0.134 & 7.41
  & 0.309 & 19.1 \\  
\hline
\label{tab:np_mult}
\end{tabular}
\end{table}

As a result, spectator neutrons reach a neutron ZDC positioned on one side of the interaction point in 73.5\% and 55\% of events, as calculated, respectively, with the WS-deformed and Bi-pyramid. Notably, a proton ZDC on the same side is more effective for identifying $^{20}$Ne--$^{20}$Ne events. Assuming its 100\% acceptance and efficiency with respect to spectator protons, the modeling predicts that 81.5\% and 66\% of collision events would be registered. Finally, it can be pointed out that 7.41\% and 19.1\% of events remain undetected by ZDCs at one side because neither neutrons nor protons are emitted in those events. Again, one can note a drastic contrast to $^{208}$Pb--$^{208}$Pb collisions at the LHC with negligible fractions of events leaving no signals in ALICE neutron ZDCs~\cite{Acharya2023,Acharya2024lumi,Acharya2025}.

\begin{figure*}[htb!]
    \centering
    \includegraphics[width=1.0\linewidth]{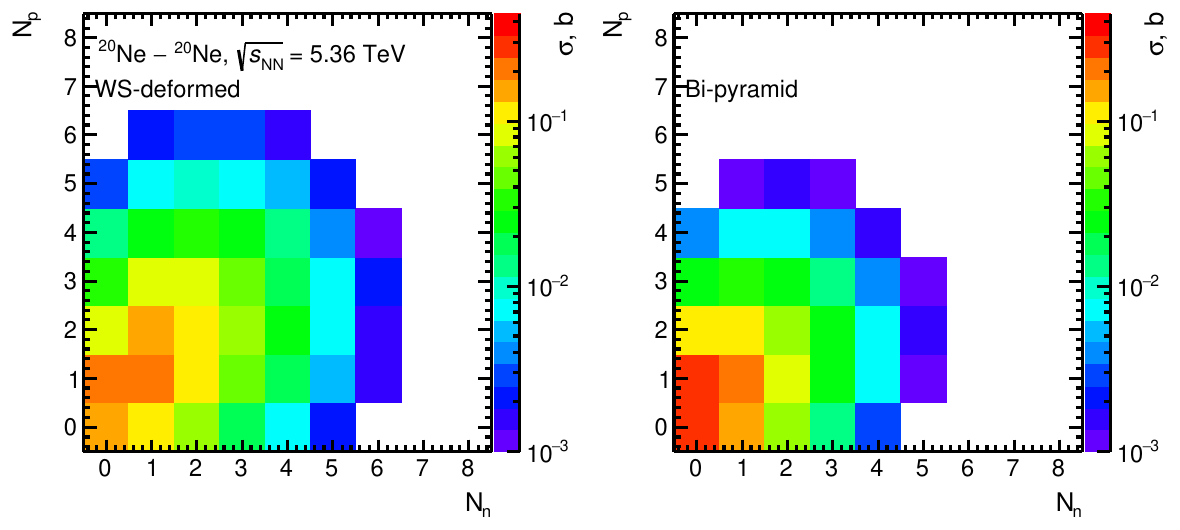}
    \caption{Cross sections of production of given numbers of spectator neutrons and protons, $\sigma(\mathrm{N_n,N_p})$ in $^{20}$Ne--$^{20}$Ne collisions at $\sqrt{s_{\rm NN}}=5.36$~TeV calculated with the WS-deformed (left) and Bi-pyramid (right) distributions of nuclear density in  $^{20}$Ne. Color scale represents the cross-sections in barns.}
    \label{fig:np_mult}
\end{figure*}

Detailed information on the cross sections of production of given numbers of spectator neutrons and protons, $\sigma(\mathrm{N_n,N_p})$, in $^{20}$Ne--$^{20}$Ne collisions at $\sqrt{s_{\rm NN}}=5.36$~TeV is presented in Fig.~\ref{fig:np_mult}. Each $(\mathrm{N_n,N_p})$ spectator channel can receive contributions from events across a range of centralities. For instance, small numbers of spectator neutrons and protons may arise in either very central or very peripheral interactions. In the former situation, only a small fraction of nucleons avoid participating in nucleon-nucleon collisions and continue propagate forward as spectators. In contrast, in the latter case many nucleons do not interact, but they stay bound in a single heavy spectator prefragment or, in particular, in several $\alpha$ clusters with sufficiently low excitation energies that inhibit additional emission of free spectator nucleons.  

As shown in Fig.~\ref{fig:np_mult}, $\sigma(\mathrm{N_n=0,N_p=1})$$>$$ \sigma(\mathrm{N_n=1,N_p=0})$, and the cross section $\sigma(\mathrm{N_n=0,N_p=1})$ exceeds those for all other $(\mathrm{N_n,N_p)}$ channels. As follows from the AAMCC-MST calculations, the emission of a single proton is frequently accompanied by $^3$H, which contains a pair of bound neutrons, as well as by $^4$He or other $2Z=A$ nuclei. This results in the inequality $\sigma(\mathrm{N_n=0,N_p>0})$$>$$ \sigma(\mathrm{N_n>0,N_p=0})$ reported in Table~\ref{tab:np_mult}.

\begin{figure*}[htb!]
    \centering
    \begin{minipage}{0.495\textwidth}
    \includegraphics[width=1.05\linewidth]{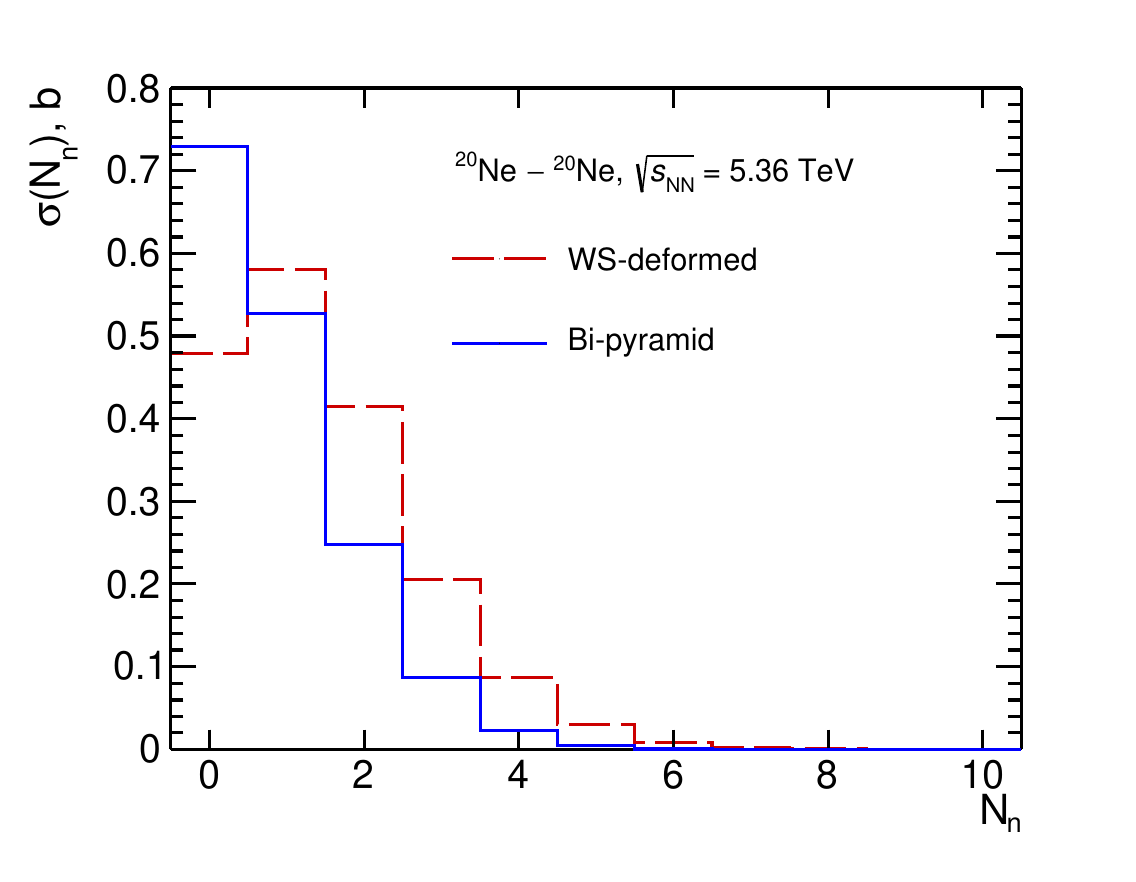}
    \end{minipage}
    \begin{minipage}{0.495\textwidth}
    \includegraphics[width=1.05\linewidth]{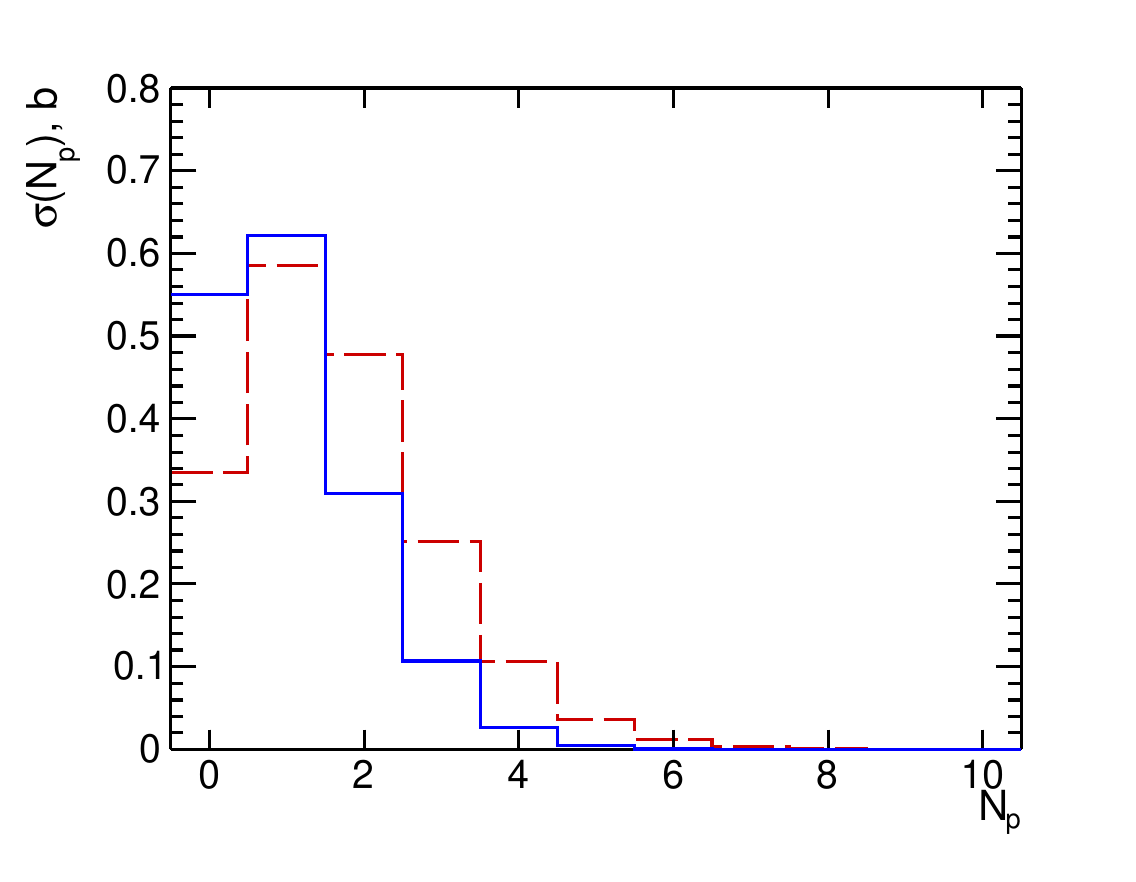}
    \end{minipage}
    \caption{Cross sections of production of given numbers of spectator neutrons, $\sigma(\mathrm{N_n})$, (left) and protons, $\sigma(\mathrm{N_p})$, (right) in $^{20}$Ne--$^{20}$Ne collisions at $\sqrt{s_{\rm NN}}=5.36$~TeV calculated with the WS-deformed and Bi-pyramid distributions of nuclear density in  $^{20}$Ne.}
    \label{fig:nucleon_mult}
\end{figure*}

The distributions of energy deposited in neutron and proton ZDCs are defined, respectively, by the multiplicity distributions of spectator neutrons and protons. The cross sections for producing a given number of neutrons, $\sigma(\mathrm{N_n})$, or a given number of protons, $\sigma(\mathrm{N_p})$, in $^{20}$Ne--$^{20}$Ne collisions at $\sqrt{s_{\rm NN}}=5.36$~TeV, obtained using the WS-deformed and Bi-pyramid distributions of nuclear density in $^{20}$Ne, are shown in Fig.~\ref{fig:nucleon_mult}. As in the $^{16}$O--$^{16}$O collisions studied in Ref.~\cite{Svetlichnyi2023}, in a sizable fraction of $^{20}$Ne--$^{20}$Ne events ($24\%$ and $45\%$) spectator neutrons are not emitted. According to the AAMCC-MST calculations with the Bi-pyramid density option, the most probable events are without neutrons, $\mathrm{N_n}=0$, see Fig.~\ref{fig:nucleon_mult}. In general, the clustered density distribution yields fewer free spectator neutrons than the WS-deformed density, for which $\mathrm{N_n}=1$ is most probable, and events with $\mathrm{N_n}=2,3,4,5$ occur much more often. 

A similar behavior is seen for $\sigma(\mathrm{N_p})$ in Fig.~\ref{fig:nucleon_mult}: the WS-deformed option yields more free spectator protons than the Bi-pyramid option, although $\mathrm{N_p}=1$ remains the dominant channel in both cases. In the Bi-pyramid representation of $^{20}$Ne, neutrons and protons are initially compactly grouped into $\alpha$ clusters, unlike in the WS-deformed case where such initial nucleon correlations are not included. As a result, when the Bi-pyramid clustering is used, the MST clustering procedure more readily incorporates nucleons into spectator fragments already at the pre-equilibrium stage rather than leaving them as unbound spectators. 

\begin{figure*}[tb!]
    \centering
    \begin{minipage}{0.495\textwidth}
    \includegraphics[width=1.05\linewidth]{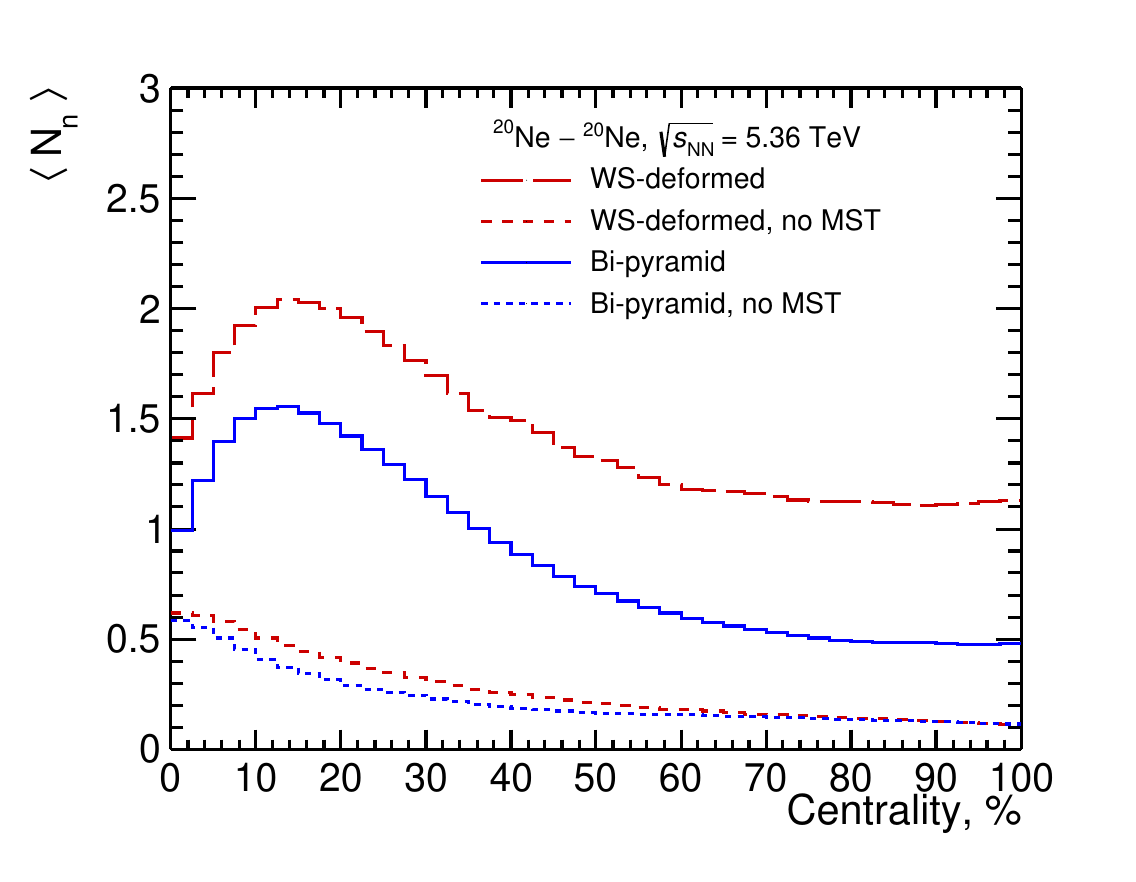}
    \end{minipage}
    \begin{minipage}{0.495\textwidth}
    \includegraphics[width=1.05\linewidth]{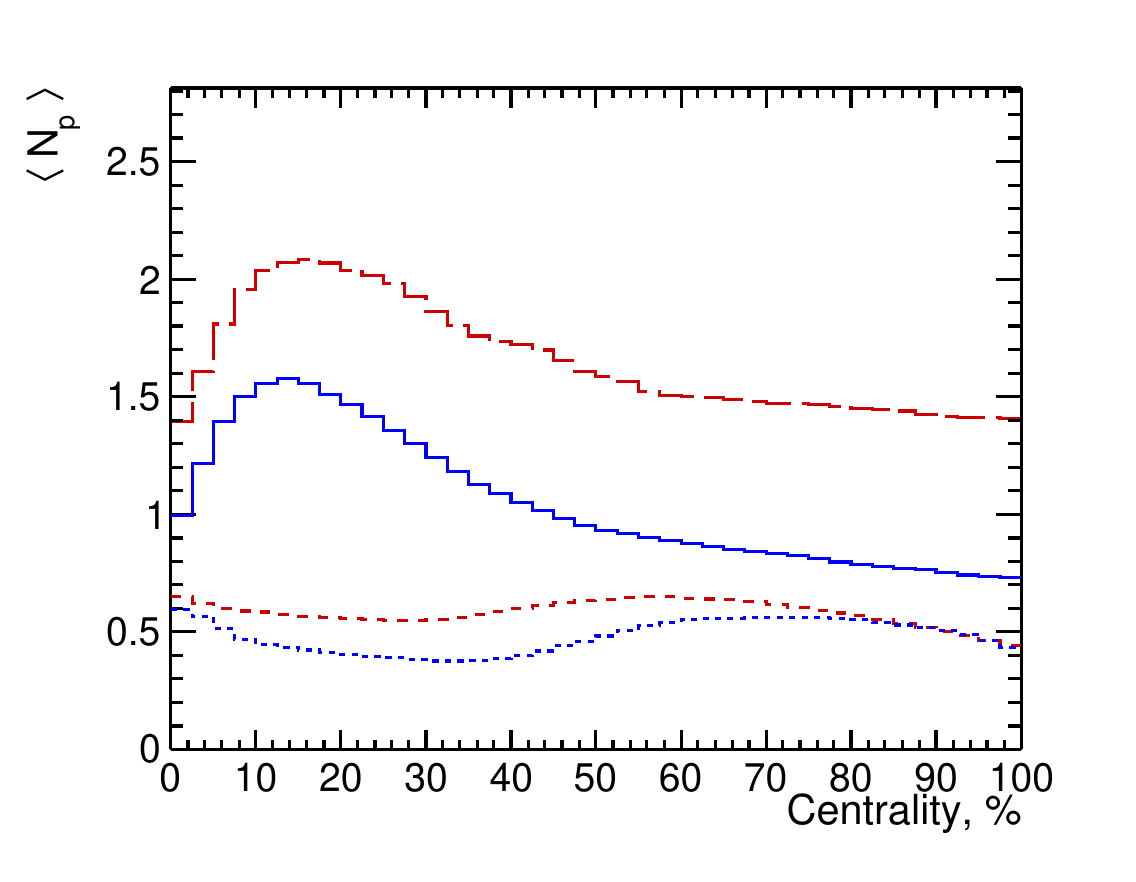}
    \end{minipage}
    \caption{Average numbers of spectator neutrons (left) and protons (right) as functions of centrality of $^{20}$Ne--$^{20}$Ne collision events at $\sqrt{s_{\rm NN}}=5.36$~TeV calculated with the WS-deformed and Bi-pyramid distributions of nuclear density in  $^{20}$Ne, with and without MST-clustering.} 
    \label{fig:N_centr}
\end{figure*}

Measuring forward neutrons with the ALICE ZDCs is one of the methods to determine the centrality of $^{208}$Pb--$^{208}$Pb collisions~\cite{Abelev2013}. It is thus worthwhile to investigate whether the same method can be applied to $^{20}$Ne--$^{20}$Ne collisions. Figure~\ref{fig:N_centr} presents the average numbers of spectator neutrons, $\langle\mathrm{N_n}\rangle$, and protons, $\langle\mathrm{N_p}\rangle$, calculated  with AAMCC-MST as functions of the $^{20}$Ne--$^{20}$Ne collision centrality at $\sqrt{s_{\rm NN}}=5.36$~TeV. As usual, the calculations were carried out with AAMCC-MST using both the WS-deformed and Bi-pyramid parameterizations of the $^{20}$Ne nuclear density. In addition, results were obtained without MST clustering.

Since fragmenting $^{20}$Ne nucleus contains equal numbers of neutrons and protons, a similarity of $\langle\mathrm{N_n}\rangle$ and $\langle\mathrm{N_p}\rangle$ obtained with each calculation option can be expected. As seen from Fig.~\ref{fig:N_centr}, this is generally true for central and semi-central collisions characterized by relatively large average numbers of free spectator neutrons and protons. However, less free nucleons are emitted as spectators in peripheral collisions because they remain bound in spectator fragments. In such collisions some excess of free protons over free neutrons is predicted, see the discussions of Table~\ref{tab:np_mult}, Figs.~\ref{fig:np_mult} and \ref{fig:nucleon_mult}.

Calculations that use the Bi-pyramid clustered structure of $^{20}$Ne yield, on average, about 1.25--2 times fewer neutrons and protons than calculations based on the WS-deformed $^{20}$Ne. This reduction of  $\langle\mathrm{N_n}\rangle$ and $\langle\mathrm{N_p}\rangle$ in the clustered $^{20}$Ne case is due to nucleons which, in particular, remain bound within $\alpha$ clusters in the final state. Such a difference naturally arises only in calculations with involving MST-cluserization. 

Omitting MST clustering in the simulations causes a pronounced decrease of the calculated  $\langle\mathrm{N_n}\rangle$ and $\langle\mathrm{N_p}\rangle$ in all centrality classes, see Fig.~\ref{fig:N_centr}. Without MST clustering, $\langle\mathrm{N_n}\rangle$ falls from around 0.5 in central collisions to about 0.1 in peripheral ones, while $\langle\mathrm{N_p}\rangle$ stays between 0.4 and 0.6 across all centrality bins. In particular, in peripheral events $\langle\mathrm{N_n}\rangle$ is markedly lower than $\langle\mathrm{N_p}\rangle$. This behavior arises because, in the AAMCC approach without MST clustering, the decay of a single excited prefragment is described by the Fermi break-up model~\cite{bondorf1995statistical}. That model yields final states consisting of various combinations of light stable nuclei, either in their ground states or in their first excited states. As these stable light nuclei have neutron-to-proton ratios slightly exceeding 1 on average, producing them requires somewhat more neutrons than protons. This accounts for the dominance of $\langle\mathrm{N_p}\rangle$ over $\langle\mathrm{N_n}\rangle$ when MST clustering is not applied, see Fig.~\ref{fig:N_centr}.

Finally, one can note a marked difference between the impact-parameter (centrality) dependence of $\langle\mathrm{N_n}\rangle$ and $\langle\mathrm{N_p}\rangle$ demonstrated in Ref.~\cite{Nepeivoda2022} for $^{208}$Pb--$^{208}$Pb collisions and the behavior presented in Fig.~\ref{fig:N_centr} for $^{20}$Ne--$^{20}$Ne. While the $^{208}$Pb--$^{208}$Pb calculations without MST clustering also give somehow smaller $\langle\mathrm{N_n}\rangle$ and $\langle\mathrm{N_p}\rangle$ than the corresponding results with MST clustering, the overall centrality trends are qualitatively different for light versus heavy colliding systems. For $^{20}$Ne--$^{20}$Ne, the dependence is relatively flat, whereas for $^{208}$Pb--$^{208}$Pb it shows a characteristic rise-and-fall pattern. In central $^{208}$Pb--$^{208}$Pb collisions, $\langle\mathrm{N_n}\rangle$ and $\langle\mathrm{N_p}\rangle$ are small because nucleons from one nucleus pass through a large $^{208}$Pb partner and therefore have a very low chance to avoid interactions and remain spectators. The values of $\langle\mathrm{N_n}\rangle$ and $\langle\mathrm{N_p}\rangle$ are likewise small in peripheral $^{208}$Pb--$^{208}$Pb events, since many nucleons remain bound in a single heavy spectator fragment and multifragmentation of the spectator matter is rare in such peripheral collisions. In contrast, because $^{20}$Ne is much thinner, the difference between central and peripheral $^{20}$Ne--$^{20}$Ne collisions is less pronounced. In addition, the excitation energy per prefragment nucleon is typically higher in peripheral $^{20}$Ne--$^{20}$Ne than in peripheral $^{208}$Pb--$^{208}$Pb. As a result, spectator nuclei with masses close to $^{20}$Ne are only seldom produced, as seen in Fig.~\ref{fig:sig_2z_eq_a}. This occurs because multifragmentation takes place not only in central but also in peripheral $^{20}$Ne--$^{20}$Ne collisions.

\section{Discussion}\label{discussion}

Treating nuclear fragmentation as a multi-stage process is a fairly common approach. It is therefore instructive to compare the AAMCC-MST model with the methods used in recent studies of $^{20}$Ne fragmentation at intermediate~\cite{Zhao2023,De2025} and high energies~\cite{Nijs2025}. 

In Refs.~\cite{Zhao2023,De2025}, the charge-changing cross sections for the fragmentation of $^{12}$C, $^{14}$N, $^{16}$O, and $^{20}$Ne projectiles with energies below 1$A$ GeV on C, Al, and Cu targets were calculated using a two-step abrasion–ablation approach. As in the AAMCC-MST model, in Refs.~\cite{Zhao2023,De2025}, the initial abrasion stage was treated within the Glauber model. In those studies, nucleus–nucleus reactions were described by assuming that the spectator matter forms a single excited residue (prefragment), which then de-excites exclusively via neutron and proton evaporation. Our approach is more general: in addition to the single spectator prefragment characteristic of peripheral nucleus–nucleus collisions, the spectator matter can alternatively be represented as a set of compact prefragments produced through the MST-clusterization. The consideration in Refs.~\cite{Zhao2023,De2025} was focused on charge-changing cross sections and the calculations were able to reproduce the measured cross sections using exclusively nucleon evaporation. Even so, it remains unclear whether a modeling of evaporation from a single excited prefragment can explain the multiple $\alpha$-particle emission seen in the data~\cite{Abd-Allah2001,Neagu2013}.  

Nuclear transmutation in $^{16}$O--$^{16}$O and $^{20}$Ne--$^{20}$Ne collisions was simulated using the Trajectum network~\cite{Nijs2025}, which also uses the GEMINI evaporation code as the only de-excitation model. Within that framework, the de-excitation of spectator matter occurs through sequential neutron and proton emission from a single spectator residue. By contrast, AAMCC-MST assumes (1) a pre-equilibrium breakup of the spectator matter into excited clusters and (2) their subsequent multifragment decays into nuclei in their ground states or in low-lying excited states. This difference results in narrower RMS spreads of the longitudinal momenta in our calculations than those reported in Ref.~\cite{Nijs2025}. In our treatment, the momenta of the initial prefragments are shared among their decay products in multifragment events. As a result, AAMCC-MST yields a smaller RMS spread compared to that in the Trajectum-plus-evaporation approach~\cite{Nijs2025}.

\section{Conclusions}\label{conc}

As demonstrated, the AAMCC-MST model provides a good overall description of the data~\cite{Abd-Allah2001,Neagu2013,Zhang2012,Zeitlin2011} on $^{20}$Ne and $^{22}$Ne fragmentation on light and heavy targets at intermediate energies.  Although the calculations largely agree with the measured charge distributions of the produced fragments, some theoretical uncertainties remain due to the choice of nuclear density profile of $^{20}$Ne (spherical, deformed, or bi-pyramid five-$\alpha$ cluster configuration) used in calculations. The model underestimates the fraction of fragmentation events with two, three, and four $\alpha$ particles with all calculation options.

The successful validation of the AAMCC-MST model with the intermediate-energy fragmentation data for $^{20}$Ne and $^{22}$Ne supports its use for calculating the production cross sections as well as transverse and longitudinal momentum distributions of $2Z=A$ spectator fragments, such as $^2$H, $^4$He, $^6$Li, $^{10}$B, $^{12}$C, $^{14}$N, and $^{16}$O, as potential contaminants in $^{20}$Ne beams circulating in the LHC. As demonstrated, the yields of $^2$H and $^4$He spectator fragments increase significantly when the Bi-pyramid $\alpha$-clustered density profile is assumed for $^{20}$Ne. Overall, across all calculation options, the model indicates that the $\alpha$ particle is the most abundant fragment among the considered $2Z=A$ species.  

On the basis of Monte Carlo modeling, a simple parametrization was proposed to describe the relative widths of the distributions of longitudinal momentum calculated per fragment nucleon for the considered $2Z=A$ fragment species. The RMS spreads of longitudinal momenta of the produced $2Z=A$ fragments rapidly decrease with the increase of $A$, from $414\cdot 10^{-4}$ for $^2$H to $74\cdot 10^{-4}$ for $^{18}$F. However, for all fragments the calculated RMS longitudinal momentum spreads were found to be much larger than that of $^{20}$Ne beams circulating in the LHC, reported as $1.13\cdot 10^{-4}$~\cite{Apollinari2015}. 

A data-driven method to estimate the contamination in LHC nucleus--nucleus collision data arising from interactions of secondary nuclei with the beam nuclei was proposed~\cite{Das2026}. This approach relies on analyzing the time dependence of the mid-rapidity particle multiplicity, but it does not quantify the relative contributions from the different contaminating secondary nuclei. By comparison, AAMCC-MST modeling offers a valuable input for detailed simulations of the transport of all possible species of $2Z=A$ fragments in the LHC. 

The AAMCC-MST can be also used to assess the performance of the Zero Degree Calorimeters (ZDCs) in $^{20}$Ne--$^{20}$Ne runs at the LHC. In ALICE, spectator neutrons and protons are measured by two pairs of neutron and proton ZDC calorimeters~\cite{Puddu2007,Gemme2009,Oppedisano2009} installed on both sides of the interaction point. As predicted by AAMCC-MST, depending on the density distribution assumed for $^{20}$Ne, between 7.41\% and 19.1\% of collision events remain undetected by ZDCs at one side, as neither neutrons nor protons are emitted in such events. AAMCC-MST simulations can also support evaluations of efficiency in detecting forward nuclear fragments with ATLAS forward detectors~\cite{Sobien2026}.

Several directions for future research can be considered. In order to validate the yields of forward neutrons and protons predicted by AAMCC-MST for $^{20}$Ne--$^{20}$Ne collisions, the respective measurements by means of ALICE ZDCs can be proposed. The modeling can be extended to the production of spectator fragments in collisions involving other light and medium-mass nuclei, such as Mg, Ar, Ca, Kr, and In. Such collision systems have recently been suggested to extend the LHC ion physics program beyond the Run~4 in future LHC operations~\cite{Waagaard2025}. While the residual background from ultraperipheral $^{16}$O--$^{16}$O collisions at the LHC, leading to electromagnetic dissociation (EMD) of $^{16}$O~\cite{Pshenichnov1998}, has been evaluated as negligible in CMS studies~\cite{Hayrapetyan2026}, the contribution of EMD in $^{20}$Ne--$^{20}$Ne collisions still needs to be thoroughly estimated, particularly using the RELDIS model~\cite{Pshenichnov1998,Pshenichnov2011}.

\bibliography{svetlichnyi_nene_lhc}%

\end{document}